\documentclass[journal,10pt]{IEEEtran}
\usepackage{mathrsfs}
\usepackage{textcomp, graphicx, subfigure}
\usepackage{epsfig}
\usepackage{pgfplots}
\usepackage{amssymb}
\usepackage{amsmath}
\usepackage{amsfonts}
\usepackage{algorithmic}
\usepackage{algorithm}
\usepackage{multirow}
\usepackage{cite}

\begin{document}
\pgfplotsset{compat=1.14}
%

\title{GRASS: \underline{GRA}ph \underline{S}pectral \underline{S}parsification Leveraging  Scalable Spectral Perturbation Analysis }


\author{Zhuo Feng,~\IEEEmembership{Senior Member,~IEEE}

\thanks{The author is  with the Department of ECE, Stevens Institute of Technology, Hoboken, NJ, 07030. Email: zfeng12@stevens.edu. This work is supported in part by  the National Science Foundation under Grants  CCF-1350206 (CAREER),  CCF-1318694 (SHF), CCF-1909105 (SHF), CCF-1618364 (SHF), and  a gift from Keysight Technologies.
%
}
}
%

\maketitle

\begin{abstract}
Spectral graph sparsification  aims to  find   ultra-sparse subgraphs  whose Laplacian matrix can well approximate the original Laplacian eigenvalues and eigenvectors.  In recent years, spectral  sparsification techniques have been extensively studied for accelerating various numerical and graph-related applications.  Prior nearly-linear-time spectral sparsification methods first extract low-stretch spanning tree from the original graph to form the backbone of the sparsifier, and then recover small portions of spectrally-critical off-tree edges to the spanning tree to significantly improve the approximation quality. However, it is not clear how many off-tree edges should be recovered for achieving a desired spectral similarity level within the sparsifier. Motivated by recent graph signal processing techniques, this paper proposes a similarity-aware spectral graph sparsification framework that leverages  efficient spectral off-tree edge embedding and filtering schemes to construct spectral sparsifiers with guaranteed spectral similarity (relative condition number) level. An iterative graph densification  scheme is also introduced to facilitate efficient and effective filtering of off-tree edges  for highly ill-conditioned problems. The proposed method has been validated using various kinds of graphs obtained from public domain sparse matrix collections relevant to VLSI CAD, finite element analysis, as well as social and data networks frequently studied in many machine learning and data mining applications. For instance, a sparse SDD matrix with $40$ million unknowns and $180$ million nonzeros can  be solved (1E-3 accuracy level) within two minutes using a single CPU core and about $6GB$ memory.
\end{abstract}

%
%


\section{Introduction}
 Spectral methods are  playing increasingly important roles in  many  graph and numerical applications \cite{teng2016scalable}, such as scientific computing \cite{spielman2014sdd}, numerical optimization \cite{christiano2011flow}, data mining \cite{peng2015partitioning}, graph analytics \cite{koren2003spectral,imre2020spectrum,zhao2018nearly},  machine learning \cite{defferrard2016convolutional,deng2019graphzoom,wang2019graspel}, graph signal processing \cite{shuman2013emerging}, and VLSI computer-aided design \cite{zhiqiang:dac17,zhuo:dac16,zhao:dac19}. For example, classical spectral graph partitioning (data clustering) algorithms embed original graphs into low-dimensional space using the first few nontrivial eigenvectors of  graph Laplacians and subsequently perform graph partitioning (data clustering) on the low-dimensional graphs to obtain  high-quality solution \cite{peng2015partitioning,wang2017towards}.  
To further push the limit of spectral methods for  large graphs, mathematics and theoretical computer science researchers have extensively studied many theoretically-sound research problems related to spectral graph theory. 
Recent \emph{{spectral graph sparsification }} research  \cite{spielman2011graph,   batson2012twice, spielman2011spectral, peng2015partitioning,cohen2017almost,Lee:2017} allows computing nearly-linear-sized subgraphs (sparsifiers) that can robustly preserve the spectra (i.e., eigenvalues and eigenvectors) of the original graph's Laplacian, which immediately leads to a series of theoretically  {{nearly-linear-time}}    numerical and graph algorithms for solving sparse  matrices, graph-based semi-supervised learning (SSL), spectral graph partitioning (data clustering), and max-flow problems \cite{miller:2010focs,  spielman2011spectral, christiano2011flow, spielman2014sdd}.  For example,  sparsified circuit networks allow for developing more scalable computer-aided (CAD) design algorithms for designing large VLSI systems \cite{zhuo:dac16,zhiqiang:dac17}; sparsified social (data) networks enable to more efficiently understand and analyze large social (data) networks \cite{teng2016scalable}; sparsified matrices can be immediately leveraged to  accelerate the solution computation  of large linear system of equations \cite{zhiqiang:iccad17}. To this end, a spectral sparsification algorithm leveraging an edge sampling scheme that sets sampling probabilities proportional to edge effective resistances (of the original graph) has been proposed in \cite{spielman2011graph}. However, it becomes a chicken-and-egg problem since calculating effective resistances (leverage scores for edge sampling) requires  solving the original graph Laplacian matrix  multiple  times (even when using Johnson–Lindenstrauss (JL) lemma \cite{spielman2011graph})  and thus can be extremely expensive for very large graphs.

  {This paper  aims to address the  standing  question whether there exists a practically-efficient, nearly-linear time spectral graph sparsification algorithm} that can immediately enable the development  of nearly-linear time sparse SDD matrix solvers and  other graph-based algorithms for large-scale, real-world  problems.   Our work is built upon the recent spectral perturbation analysis framework that allows for  highly-scalable spectral  sparsification of large (weighted) undirected graphs \cite{zhuo:dac16,zhuo:dac18}, which also shows connections with the Courant-Fishcher theorem for generalized eigenvalues as well as recent graph signal processing techniques \cite{shuman2013emerging}. 
  
  Our method starts by extracting a spectrally-critical spanning tree subgraph as a backbone of the sparsifier, and subsequently recovers a small portion  spectrally-critical off-tree edges to the spanning tree. In many scientific computing and graph-related applications, it is usually desired to construct    sparsifiers according to a given spectral similarity level:  introducing too few edges may lead to poor approximation of the original graph, whereas too many edges can result in rather high computational complexity. For example, when using a preconditioned conjugate gradient (PCG) method  to solve a symmetric diagonally dominant (SDD) matrix for multiple right-hand-side (RHS) vectors, it is hoped the PCG solver would converge to a good solution as fast as possible,   which usually requires the sparsifier (preconditioner) to be highly spectrally-similar to the original problem or achieve relatively a small condition number; on the other hand, in many graph partitioning tasks, only the Fiedler vector (the first nontrivial eigenvector) of graph Laplacian is needed \cite{spielmat1996spectral}, so even a  sparsifier with a much lower spectral similarity should suffice. 
To this end,   this work introduces a similarity-aware spectral graph sparsification framework that leverages efficient  spectral off-tree edge embedding and filtering schemes  to construct spectral sparsifiers with guaranteed spectral similarity.    The   contribution of this work has been summarized as follows: 
\begin{enumerate}
  \item We present a  nearly-linear time yet practically-efficient framework for constructing ultra-sparsifier subgraph from a  spanning tree subgraph via efficient spectral perturbation analysis of generalized eigenvalue problem. The proposed algorithm allows effectively fixing the largest generalized eigenvalues, by recovering the most spectrally-critical off-tree edges to the  spanning-tree subgraph.
  
\item We present a similarity-aware spectral  sparsification framework by leveraging   scalable   spectral off-tree edge embedding and filtering schemes motivated by recent graph signal processing techniques \cite{shuman2013emerging}. Such a scheme enables to flexibly trade off the complexity and spectral similarity of the sparsified graph.

 \item For highly ill-conditioned problems, we introduce an iterative   graph densification scheme  to more effectively fix the most problematic eigenvalues by progressively improving the ultra-sparsifier subgraphs.  Compared to the single-pass spectral sparsification scheme, the proposed scheme can always achieve greater reduction of  relative condition numbers, thereby leading to more effective spectral approximation of the original graph Laplacian.
    \item Extensive experiments have been conducted to validate   the proposed method in various numerical and graph-related applications, such as solving sparse SDD matrices,  spectral graph partitioning, as well as compression of large social and data  networks. Very promising  results  for even extremely ill-conditioned VLSI,   thermal and finite element analysis problems have been obtained.
\end{enumerate}

The rest of this paper is organized as follows.  Section
\ref{background_sec} provides a brief introduction to graph Laplacians  and the state of the art in spectral sparsification of graphs. In Sections \ref{main_sec} and \ref{sec:similarity}, a scalable similarity-aware spectral sparsification method based on spectral perturbation analysis  is described in detail.  Section
\ref{result_sec}  demonstrates extensive experimental results for a variety of real-world, large-scale sparse Laplacian matrix and  graph problems, which is followed by the conclusion of this work in Section \ref{conclusion}.

\section{Background}\label{background_sec}
\subsection{Graph Laplacian Matrices and Quadratic Forms}
\begin{figure}
\centering \epsfig{file=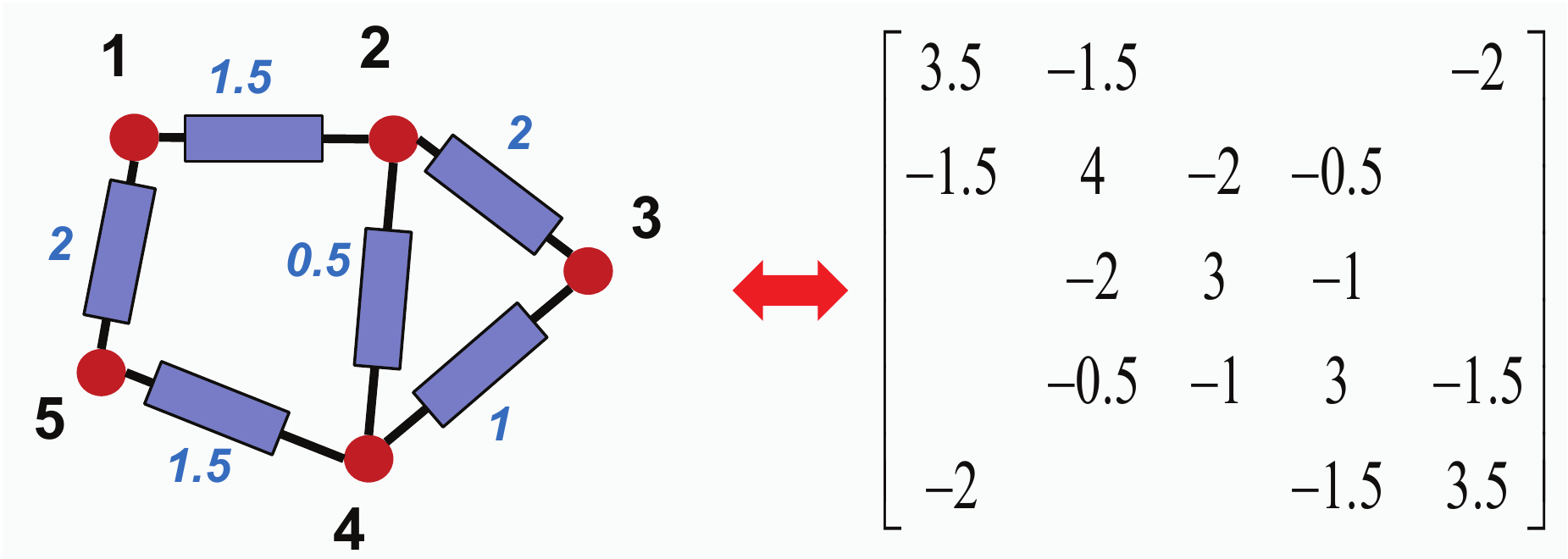, scale=0.458 } \caption{A resistor network (conductance value of each element is shown) and its graph Laplacian matrix. \protect\label{fig:graph_laplacian}}
\end{figure}

Consider a weighted, undirected  and connected graph   $G=(V,E,\omega)$, where
$V$ denotes a set of vertices, $E$ denotes a set of edges, and $\omega$ denotes a weight
function that assigns a positive weight to each edge. As shown in Fig. \ref{fig:graph_laplacian}, the Laplacian matrix of graph $G$ can be defined as  follows:
\begin{equation}\label{formula_laplacian}
\mathbf{L_G}(p,q)=\begin{cases}
-\omega(p,q) & \text{ if } (p,q)\in E \\
\sum\limits_{(p,t)\in E}\omega(p,t) & \text{ if } (p=q) \\
0 & \text{ if } otherwise.
\end{cases}
\end{equation}
It can be shown that  every graph Laplacian  matrix is an SDD matrix, which also can  be considered as an admittance matrix of a resistor circuit network. For any  real vector $\mathbf{x}\in {\mathbb{R} ^V}$, the Laplacian quadratic form of graph $G$ is defined as:
 \begin{equation}\label{formula_laplacian_form}
\mathbf{{x^\top}L_G x} = \sum\limits_{\left( {p,q} \right) \in E}
{{\omega_{p,q}}{{\left( {x\left( p \right) - x\left( q \right)}
\right)}^2}},
\end{equation}
which can be considered as  the Joule heat dissipated in the resistor network when $\mathbf{x}$ is a voltage vector.

 \subsection{Graph Sparsification and Its Applications}
Classic graph sparsification problem can be described as follows: given a graph $G=(V,E,\omega)$ and its  Laplacian matrix $\mathbf{L_G}$, graph sparsification aims to find a subgraph (a.k.a graph sparsifier)  $P=(V,E_s,\omega_s)$ and its   Laplacian  $\mathbf{L_P}$ so that the sparse subgraph will retain  all vertices but significantly less number of edges compared to the original graph.
\begin{figure}
\centering \epsfig{file=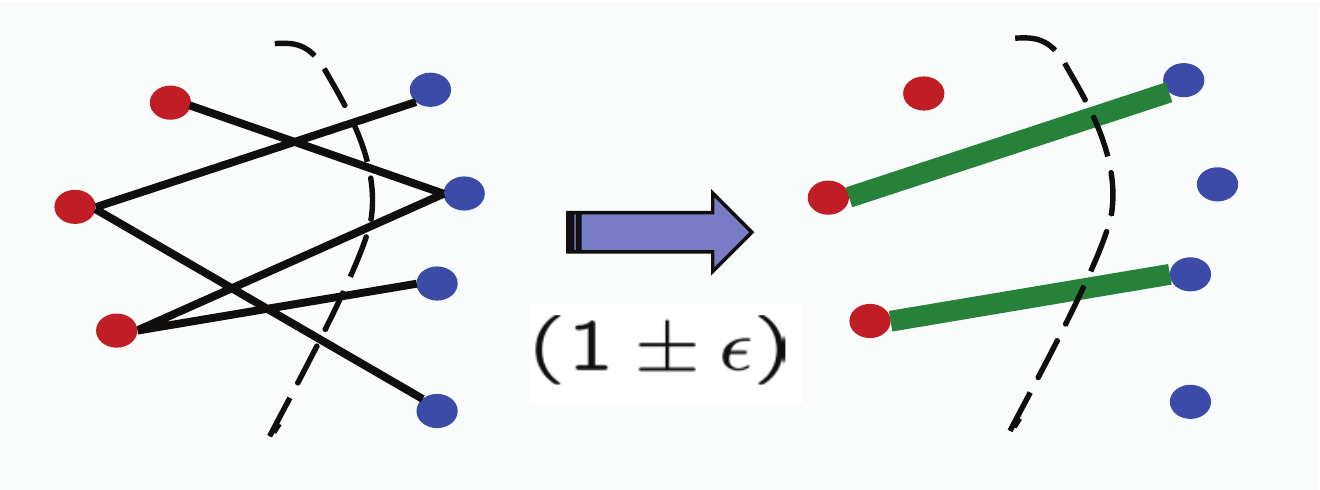, scale=0.65 } \caption{Cut sparsifier preserves cuts of a graph. \protect\label{fig:cut_sparsifier}}
\end{figure}

Graph sparsifiers typically fall into the following two categories: the cut sparsifier \cite{benczur1996approximating} and spectral sparsifier \cite{spielman2011spectral}.  The cut sparsifier preserves the values of cuts in a graph as shown in Fig. \ref{fig:cut_sparsifier}, whereas   {spectral  sparsifier}  preserves eigenvalues and eigenvectors  of the original graph. It has been shown that a good cut sparsifier may not always be a good spectral sparsifier, while the spectral sparsifier is always  a stronger notion than the cut sparsifier \cite{spielman2011spectral}.

To  illustrate the role of   graph sparsification   in numerical computation applications, consider the following example of the standard PCG algorithm for solving SDD matrices.  PCG can find an $\epsilon$-accurate solution in at most $O(\kappa(\mathbf{{L_G},{L_P}})^{1/2}\log \epsilon^{-1})$ iterations, where the relative condition number $\kappa(\mathbf{{L_G},{L_P}})$  is defined as follows:
\begin{equation}\label{formula_condition_number}
\kappa(\mathbf{{L_G},{L_P}})=\frac {\lambda_{max}}
{\lambda_{min}},
\end{equation}
where $\lambda_{min}$ and $\lambda_{max}$ denote the smallest and largest nonzero generalized eigenvalues\footnote{The smallest eigenvalue of a Laplacian matrix is always $0$ with the corresponding all-1s eigenvector. For a disconnected graph, the number of zero eigenvalues equal to the number of disconnected components, which is equal to the algebraic multiplicity of $0$ in the Laplacian.} that satisfy:
\begin{equation}\label{formula_generalized_eigen}
\mathbf{L_G u=\lambda L_P u},
\end{equation}
with $\mathbf{u}$ denoting the eigenvector corresponding to the generalized eigenvalue $\lambda$. It is desired that  the preconditioner $\mathbf{L_P}$ matrix should be very similar to $\mathbf{L_G}$ and result in a much smaller relative condition number. It is also desired that $\mathbf{L_P}$ should be much sparser   than  $\mathbf{L_G}$   so that the overall cost of the preconditioned iterations can be much lower than directly solving the original  matrix. It is obvious that a graph sparsifier with good spectral approximation of the original matrix can be immediately adopted as a preconditioner for solving SDD or SDD-like matrices \cite{miller:2010focs, spielman2014sdd, xueqian:iccad12,xueqian:dac12,xueqian:tcad15,lengfei:tcad15,lengfei:iccad13}.

 \subsection{Spectral Graph Sparsification}
   \begin{figure}
\centering \epsfig{file=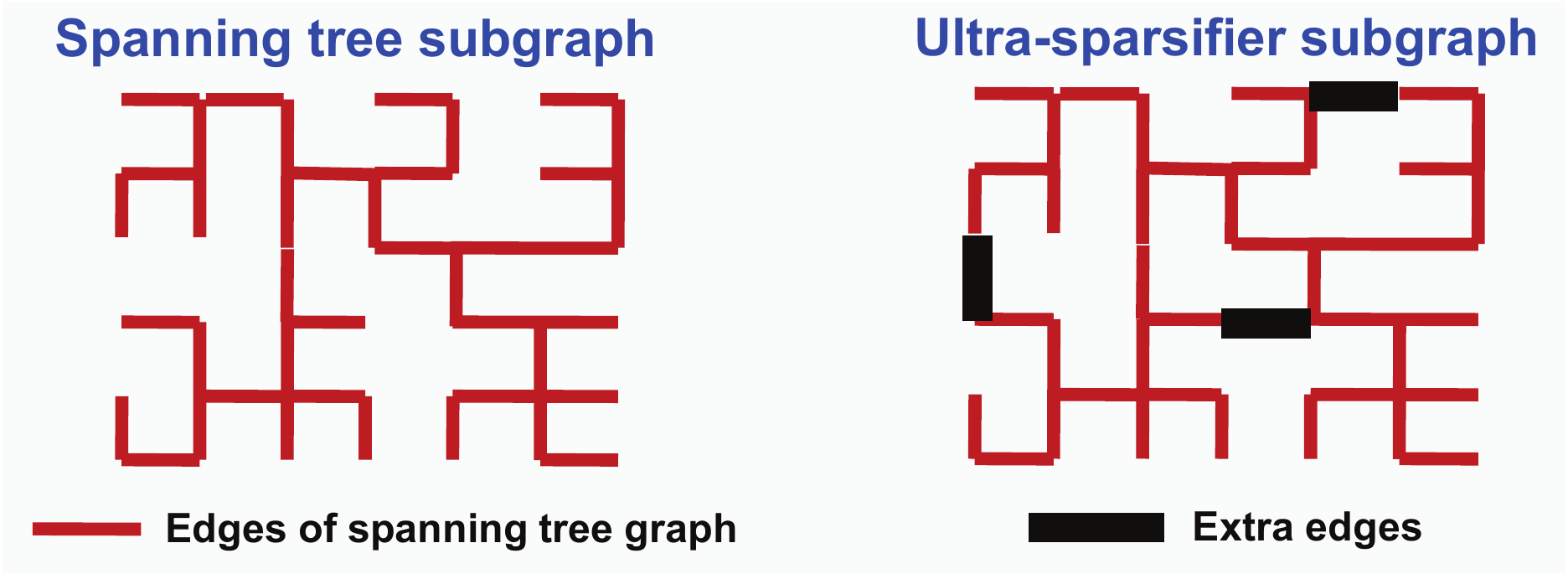, scale=0.45 } \caption{A spanning tree and its ultra-sparsifier subgraph. \protect\label{fig:ultrasparsifier}}
\end{figure}
 Graphs $G$ and $P$ are said to be $\sigma-$spectrally similar if for all real vectors $\mathbf{x} \in \mathbb{R}^V$ the following holds \cite{batson2013spectral}:
 \begin{equation}\label{formula_spectral_similar}
\mathbf{\frac{x^\top{L_P}x}{\sigma}\le x^\top{L_G}x \le \sigma x^\top{L_P}x}.
\end{equation}
  It can be shown that the relative condition number $\kappa(\mathbf{{L_G},{L_P}})\le\sigma^2$. Note that   a graph sparsifier  that can result in a smaller relative condition number will also be spectrally similar to the original graph,  and thus can lead to faster convergence of iterative methods, such as Krylov-subspace iterative methods. For complete graphs the Ramanujan graphs are among the best spectral graph sparsifiers \cite{batson2012twice}, whereas for arbitrary graphs the linear-sized Twice-Ramanujan graphs can achieve the same spectral similarity with  twice as many as the edges in Ramanujan graphs \cite{Lee:2017,batson2012twice}. However, it still remains unclear if there is a practically-efficient algorithm for constructing  linear-sized spectral sparsifiers.
  
In this work, we define  \emph{spectrally-critical} edges to be the ones that can mostly perturb the graph spectral  properties, such as the first few Laplacian eigenvalues and eigenvectors. Recent approaches for constructing nearly-linear-sized spectral sparsifiers  aim to dramatically reduce relative condition number, which typically include the following two key steps (as shown in Fig. \ref{fig:ultrasparsifier}) \cite{ spielman2011graph,Kolla:2010stoc,miller:2010focs,spielman2014sdd}:
  \begin{enumerate}
    \item Extract an initial spanning tree  from the original graph as a backbone of the sparsifier;
    \item Recover a small number of  spectrally-critical off-tree edges    to the spanning tree  to form an ultra-sparsifier subgraph.
  \end{enumerate}
Note that since  the smallest   generalized eigenvalue $\lambda_{min}$ of a spanning tree subgraph (without scaling) is no smaller than $1$,  we always have $\lambda_{max}\ge\kappa(\mathbf{{L_G},{L_P}})=\frac{\lambda_{max}}{\lambda_{min}}$.

For step 1), recent theoretical computer science research results suggest that low-stretch spanning trees should  be constructed since they will result in an upper bound of $\lambda_{max}<O(m\log n(\log\log n)^2) $ that can immediately lead to the development of nearly-linear time algorithms for solving SDD matrices \cite{miller:2010focs}, where $m$ denotes the number of nonzeros and $n$ the number of equations in the matrix. Towards this goal, nearly-linear time low-stretch spanning tree algorithms based on star- and petal-decomposition methods have been proposed \cite{elkin2008lst,abraham2012}.

For step 2), it requires to recover the most spectrally-critical off-tree edges to the spanning tree for constructing the ultra-sparsifier, thereby drastically improving   spectral approximation of the sparsifier. To this end, effective-resistance based edge sampling scheme \cite{spielman2011graph} can be adopted for recovering these off-tree edges.  However,  calculating effective resistances (leverage scores for edge sampling) requires  solving  graph Laplacian   multiple  times (even when using the Johnson-Lindenstrauss Lemma \cite{spielman2011graph})  and thus can be rather expensive for very large graphs.  It is also suggested to use stretch to replace effective resistance for edge sampling, which is more computationally efficient but will increase the number of edges sampled \cite{miller:2010focs}.

%


\section{Spectral Graph Sparsification Via Efficient Spectral Perturbation Analysis}\label{main_sec}
In this paper, we introduce a practically-efficient, nearly-linear time spectral graph sparsification algorithm that can be   efficiently applied to sparsify  large-scale real-world graphs (Laplacian matrices). We show that for an initial spectrally-critical spanning-tree subgraph, such as a low-stretch spanning tree, the proposed algorithm can always efficiently  identify the most spectrally-critical off-tree edges to be added to the spanning tree, thereby drastically reducing the relative condition number  of the preconditioned system. As a result, an ultra-sparse yet spectrally-similar subgraph can be constructed  and leveraged for solving sparse SDD matrices as well as other graph related  problems in nearly-linear time.

\subsection{Overview of Our Approach}
The overview of the proposed method for similarity-aware spectral sparsification of undirected graphs has been summarized as follows. For a given input graph, the following key procedures are involved in the proposed algorithm flow: \textbf{(a)}  low-stretch spanning tree \cite{elkin2008lst,abraham2012}  extraction based on its original  graph Laplacian; \textbf{(b)} spectral (generalized eigenvalue)  embedding and filtering of off-tree edges by leveraging the  recent spectral  perturbation analysis framework \cite{zhuo:dac16}; \textbf{(c)} iterative sparsifier improvement (graph densification) by gradually adding small portions of dissimilar off-tree edges  to the spanning tree.

In the rest of this paper,  we assume that $G=(V,E,w)$ is a weighted, undirected and connected graph, whereas $P=(V,E_s,w_s)$ is its sparsifier. To simplify the our analysis, we assume the edge weights in the sparsifier remain the same as the original ones, though  iterative edge  weight re-scaling schemes via stochastic gradient descent (SGD) optimization \cite{zhao:dac19}  can be applied to further improve the spectral approximation. The descending   eigenvalues of $\mathbf{L^+_P L_G}$ are denoted by $\lambda _{max}={\lambda _1} \ge {\lambda _2} \ge  \cdots  \ge {\lambda _n} \ge 1$, where $\mathbf{L^+_P}$ denotes the  Moore-Penrose  pseudoinverse of $\mathbf{L_P}$.

\subsection{Spanning Tree as A Spectral Sparsifier}
\begin{figure}
\centering \epsfig{file=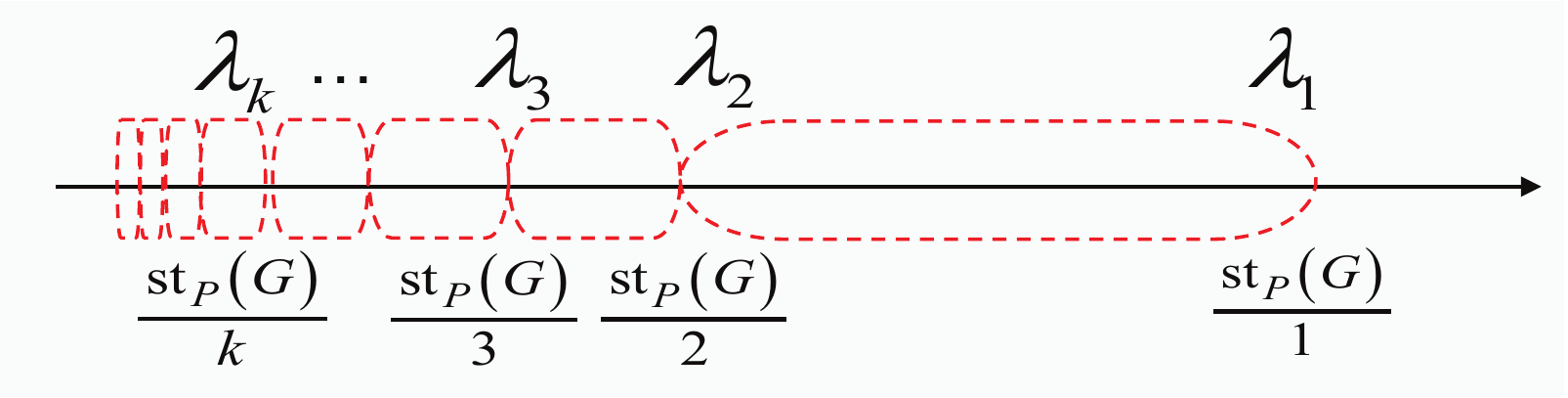, scale=0.535 } \caption{A nearly worst-case distribution of generalized eigenvalues for a spanning-tree preconditioned system. \protect\label{fig:eigworst}}
\end{figure}
Recent  work shows that when using a spanning tree subgraph  as a spectral graph sparsifier or preconditioner, $\mathbf{{L^+_P}L_G}$ will not have many large eigenvalues \cite{spielman2009note}: it has at most $k$ eigenvalues greater than $\frac {\textstyle{{{\rm{s}}{{\rm{t}}_{P}}\left( G \right)}}} {k} $, where ${{\rm{s}}{{\rm{t}}_{P}}\left( G \right)} $ denotes the stretch of the original graph $G$ with respect to the spanning tree subgraph $P$  defined as \cite{spielman2009note}:
\begin{equation}\label{formula_stretch_graph}
{\rm{s}}{{\rm{t}}_P}\left( G \right) = \sum\limits_{\left( {p,q} \right) \in E } {{\rm{s}}{{\rm{t}}_P}\left( {p,q} \right)},
\end{equation}
where ${{\rm{s}}{{\rm{t}}_{P}}\left( {p,q} \right)} $ denotes the stretch of    {  edge} $\left( {p,q} \right)$ defined as:
\begin{equation}\label{formula_stretch_edge}
{\rm{s}}{{\rm{t}}_P}\left( {p,q} \right) = {\omega_{p,q}}\left( {\sum\limits_{f \in S} {\frac{1}{{{\omega_f}}}} } \right),
\end{equation}
where $S$ denotes the set of edges in the path in the spanning tree $P$ from $p$ to $q$.  A nearly-worst case distribution of eigenvalues with $\lambda _{i}\le{{\rm{s}}{{\rm{t}}_{P}}\left( G \right)} /i$ has been shown in Fig. \ref{fig:eigworst}.
\begin{figure}
\centering \epsfig{file=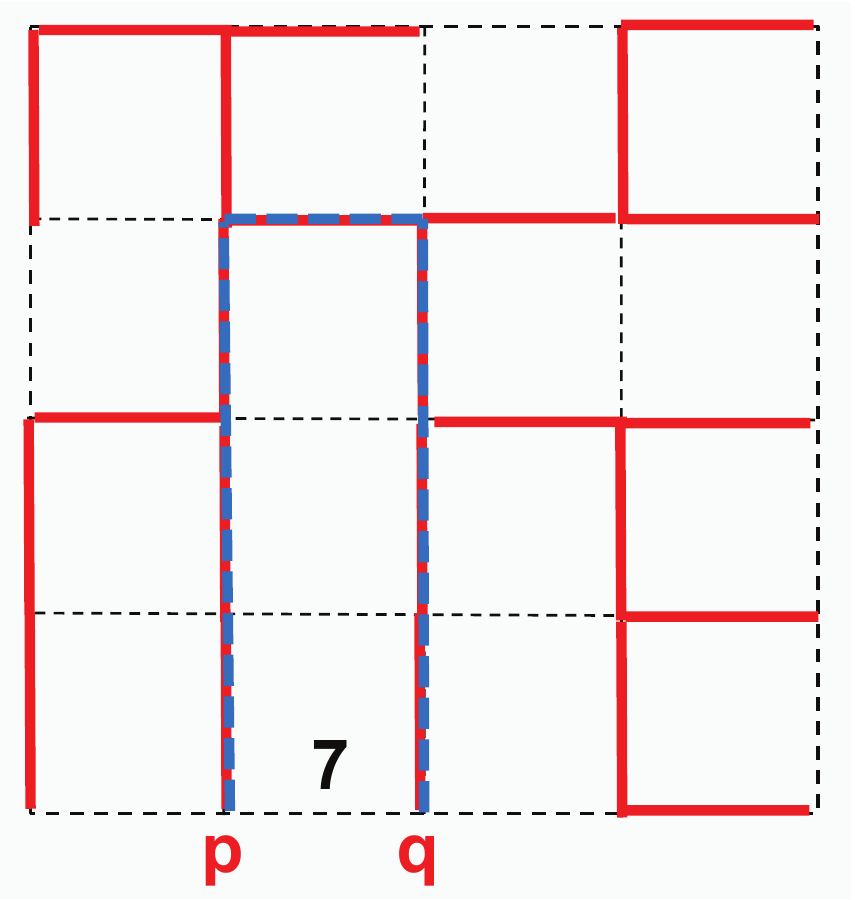, scale=0.64} \caption{The stretch of an off-tree edge $(p,q)$ is computed by ${{\rm{s}}{{\rm{t}}_{P}}\left( {p,q} \right)}=7 $ for a weighted graph with equal edge  weight. \protect\label{fig:lowstretch}}
\end{figure}


Define $\mathbf{{e_{p}}}\in {\mathbb{R}^V}$ to be a vector with  only the $p$-th element being $1$ and others being $0$. Also define  $\mathbf{{e_{p,q}=e_{p}-e_{q}}}$. Then the trace of $\mathbf{{L^+_P}L_G}$ becomes \cite{spielman2009note}:
\begin{equation}\label{formula_trace}
\begin{split}
{\rm{Tr}}\left( \mathbf{{{L^+_P}L_G}} \right) &=\sum\limits_{i = 1}^n \lambda_i  = \sum\limits_{\left( {p,q} \right) \in E } {{\omega _{p,q}}} {\rm{Tr}}\left( \mathbf{{{L^+_P} e_{p,q} {{e^\top_{p,q}}}}} \right)\\
 &= \sum\limits_{\left( {p,q} \right) \in E } {{\omega _{p,q}}} {\rm{Tr}}\left( \mathbf{{{{e^\top_{p,q}}}{L^+_P}e_{p,q}}} \right)\\
 &= \sum\limits_{\left( {p,q} \right) \in E } {{\omega _{p,q}}} \mathbf{{e_{p,q}^\top}{L^+_P}e_{p,q}}\\&={{\rm{s}}{{\rm{t}}_{P}}\left( G \right)}\ge\lambda_1=\lambda_{max}.
\end{split}
\end{equation}
The above result indicates that the total stretch ${\rm{s}}{{\rm{t}}_{P}}\left( G \right) $ is bounded by $\lambda_{1}$. Therefore, it is suggested that a spanning tree with low stretch should be constructed so that the preconditioned system will have a  small relative condition number. It has been shown that if each of the largest eigenvalues   can be fixed (dramatically decreased) by recovering a small number of  off-tree edges to the spanning tree, an ultra-sparsifier with totally $n+o(n)$ edges can be created to provide a good spectral approximation of  $G$  \cite{Kolla:2010stoc,miller:2010focs,spielman2014sdd}. For instance, it has been shown that an ultra-sparsifier with a relative condition number $\alpha k^2$ can be constructed by adding at most $n/k$  off-tree edges to the spanning tree subgraph \cite{spielman2014sdd}. 

\subsection{Perturbation Analysis of Generalized Eigenvalue Problems}
Consider the following first-order eigenvalue perturbation problem:
\begin{equation}\label{formula_eig_perturb1}
\mathbf{L_G}\left( \mathbf{{{u_i} + \delta {u_i}}} \right) = \left( \mathbf{{{\lambda _i} + \delta {\lambda _i}}} \right)\left( \mathbf{{L_P + \delta L_P}} \right)\left( \mathbf{{{u_i} + \delta {u_i}}} \right),
\end{equation}
where a perturbation $\delta \mathbf{L_P}$  is applied to $\mathbf{L_P}$, leading to the perturbations in generalized eigenvalues and eigenvectors  ${\lambda _i} + \delta {\lambda _i}$ and $\mathbf{{u_i} + \delta {u_i}}$ for $i=1,...,n$, respectively. By keeping only the first-order terms, (\ref{formula_eig_perturb1}) becomes:
\begin{equation}\label{formula_eig_perturb2}
\mathbf{L_G\delta {u_i} = {\lambda _i}L_P\delta {u_i} + \delta {\lambda _i}L_P{u_i} + {\lambda _i}\delta L_P{u_i}}.
\end{equation}
Expressing $\mathbf{\delta {u_i}}$ with the original eigenvectors $\mathbf{{u_j}}$ leads to:
\begin{equation}\label{formula_eigvec}
\delta \mathbf{{u_i} = \sum\limits_{j = 1}^n {{\zeta _{ij}}{u_j}}},
\end{equation}
where $\mathbf{{u_j}}$ can always be constructed to satisfy:
\begin{equation}\label{formula_p-orth}
\mathbf{u_i^\top L_Pu_j^{}} = \left\{ \begin{array}{l}
1,i = j\\
0,i \ne j.
\end{array} \right.\end{equation}
Substituting (\ref{formula_eigvec}) into (\ref{formula_eig_perturb2}) leads to:
\begin{equation}\label{formula_eig_perturb3}
\begin{array}{l}
  {\sum\limits_{j = 1}^n \mathbf{{{\zeta _{ij}}{\lambda _j}L_P{u_j}}} }  \\= {\lambda _i}\mathbf{L_P\left( {\sum\limits_{j = 1}^n {{\zeta _{ij}}{u_j}} } \right) + \delta {\lambda _i}L_P{u_i} + {\lambda _i}\delta L_P{u_i}}.
\end{array}
\end{equation}
Multiplying $u_i^\top$ to both sides of (\ref{formula_eig_perturb3}) results in:
\begin{equation}\label{formula_eig_perturb4}
\begin{array}{l}
\mathbf{u_i^\top\left( {\sum\limits_{j = 1}^n {{\zeta _{ij}}{\lambda _j}L_P{u_j}} } \right)} = \\
\mathbf{{\lambda _i}u_i^\top L_P\left( {\sum\limits_{j = 1}^n {{\zeta _{ij}}{u_j}} } \right) + \delta {\lambda _i}u_i^\top L_P{u_i} + {\lambda _i}u_i^\top\delta L_P{u_i}},
\end{array}
\end{equation}
which immediately leads to:
\begin{equation}\label{formula_eig_perturb5}
\mathbf{\delta {\lambda _i} =  - {\lambda _i}\frac{{u_i^\top\delta L_P{u_i}}}{{u_i^\top L_P{u_i}}} =  - {\lambda _i}u_i^\top\delta L_P{u_i}}.
\end{equation}
Expanding $\delta \mathbf{L_P}$ that includes multiple  extra off-tree edges $(p,q)$ leads to:
\begin{equation}\label{formula_deltaP}
\mathbf{\delta \mathbf{L_P = \sum\limits_{(p,q)\in E\setminus E_s}^{} {{\omega_{p,q}}e_{p,q}e_{p,q}^\top}}}.
\end{equation}
  The above leads to the following based on (\ref{formula_eig_perturb5}):
\begin{equation}\label{formula_eig_perturb6}
\mathbf{\delta {\lambda _i} =  - {\lambda _i}\sum\limits_{(p,q)\in E\setminus E_s}^{} {{w_{p,q}}u_i^\top e_{p,q}e_{p,q}^\top}{u_i}}.
\end{equation}

It is obvious from (\ref{formula_eig_perturb6})  that  the reduction of $\lambda _i$ is  proportional to the Joule heat produced by the extra off-tree edges  when its unperturbed eigenvector $\mathbf{u_i}$ is applied as a node-voltage vector.
For instance,  the voltage difference between nodes $p$ and $q$ is computed by
\begin{equation}\label{formula_vol_diff}
v_{p,q}=\mathbf{{u_i^\top {e_{p,q}} }},
\end{equation}
 which results in the following Joule heat for the off-tree edge $(p,q)$:
  \begin{equation}\label{formula_Joule}
h_{p,q}={\omega_{p,q}}v^2_{p,q}.
\end{equation}
Therefore, the perturbation of  eigenvalue $\lambda _i$ becomes:
    \begin{equation}\label{formula_eig_perturb7}
\delta {\lambda _i}=-\lambda _i \sum\limits_{(p,q)\in E\setminus E_s}^{} h_{p,q}.
\end{equation}
 Consequently,   adding the off-tree edges with the largest Joule heat values computed using the dominant generalized eigenvector ($\mathbf{u_1}$) that corresponds to the largest eigenvalue ($\lambda_1$)  will dramatically reduce  the relative condition number thereby improving the spectral approximation of the subgraph. Repeating the above procedures for all large eigenvalues will lead to  very good spectral graph sparsifiers.

 \subsection{Why  Dominant Generalized  Eigenvectors?}
As shown in the previous perturbation analysis,  dominant generalized eigenvectors can help identify the  most spectrally-critical off-tree edges. Alternatively,  the following \textbf{Courant-Fischer theorem } can be leveraged for better understanding why   generalized eigenvectors should be used for spectral sparsification tasks.
Assigning each node in the graph with an integer value either $0$ or $1$, the corresponding Laplacian  quadratic form measures the boundary size (cut) of a node set. For example, if a node set $Q$ is defined as
\begin{equation}
Q \overset{\mathrm{def}}{=}\left\{ q \in V: x(q)=1 \right\},
\end{equation}
then the number of edges going out of   $Q$ becomes:
\begin{equation}
\mathbf{x}^\top \mathbf{L}_G \mathbf{x}=cut(Q,  \overline{Q} )= |\partial_G (Q)|, 
\end{equation}
where the boundary of $Q$ in $G$ is defined as
\begin{equation}
\partial_G (Q) \overset{\mathrm{def}}{=}\left\{ (p,q)\in E: p\notin  Q, q \in Q\right\}.
\end{equation} 
The {Courant-Fischer theorem } for generalized eigenvalue problems allows finding dominant eigenvalues and eigenvectors by solving the following optimization task:
\begin{equation}\label{formula_courant-fischer-max}
\lambda_{max}=\mathop{\max_{|\mathbf{x}|\neq 0}}_{\mathbf{x}^\top\mathbf{1}=0}\frac{\mathbf{x}^\top \mathbf{L}_G \mathbf{x}}{\mathbf{x}^\top \mathbf{L}_P \mathbf{x}}\ge\mathop{\max_{|\mathbf{x}|\neq 0}}_{x(p)\in \left\{0,1 \right\}}\frac{\mathbf{x}^\top \mathbf{L}_G \mathbf{x}}{\mathbf{x}^\top \mathbf{L}_P \mathbf{x}}=\mathop{\max}\frac{|\partial_G (Q)|}{|\partial_P (Q)|},
\end{equation}
where  $\mathbf{1}\in {\mathbb{R}^V}$ is the all-one vector,
and   the boundary of $Q$ in $P$ is defined as
\begin{equation}
\partial_P (Q) \overset{\mathrm{def}}{=}\left\{ (p,q)\in E_s: p\notin  Q, q \in Q\right\},
\end{equation} 
which leads to the following equations:
\begin{equation}
\mathbf{x}^\top \mathbf{L}_G \mathbf{x}=|\partial_G (Q)|,\mathbf{x}^\top \mathbf{L}_P \mathbf{x}=|\partial_P (Q)|.
\end{equation}
 Then (\ref{formula_courant-fischer-max}) indicates that  finding the dominant generalized eigenvector would be closely related to finding $Q$ such that $\frac{|\partial_G (Q)|}{|\partial_P(Q)|}$ or the ratio of the boundary  sizes in the original graph $G$ and subgraph $P$   is maximized.  
 \begin{figure}
\centering \epsfig{file=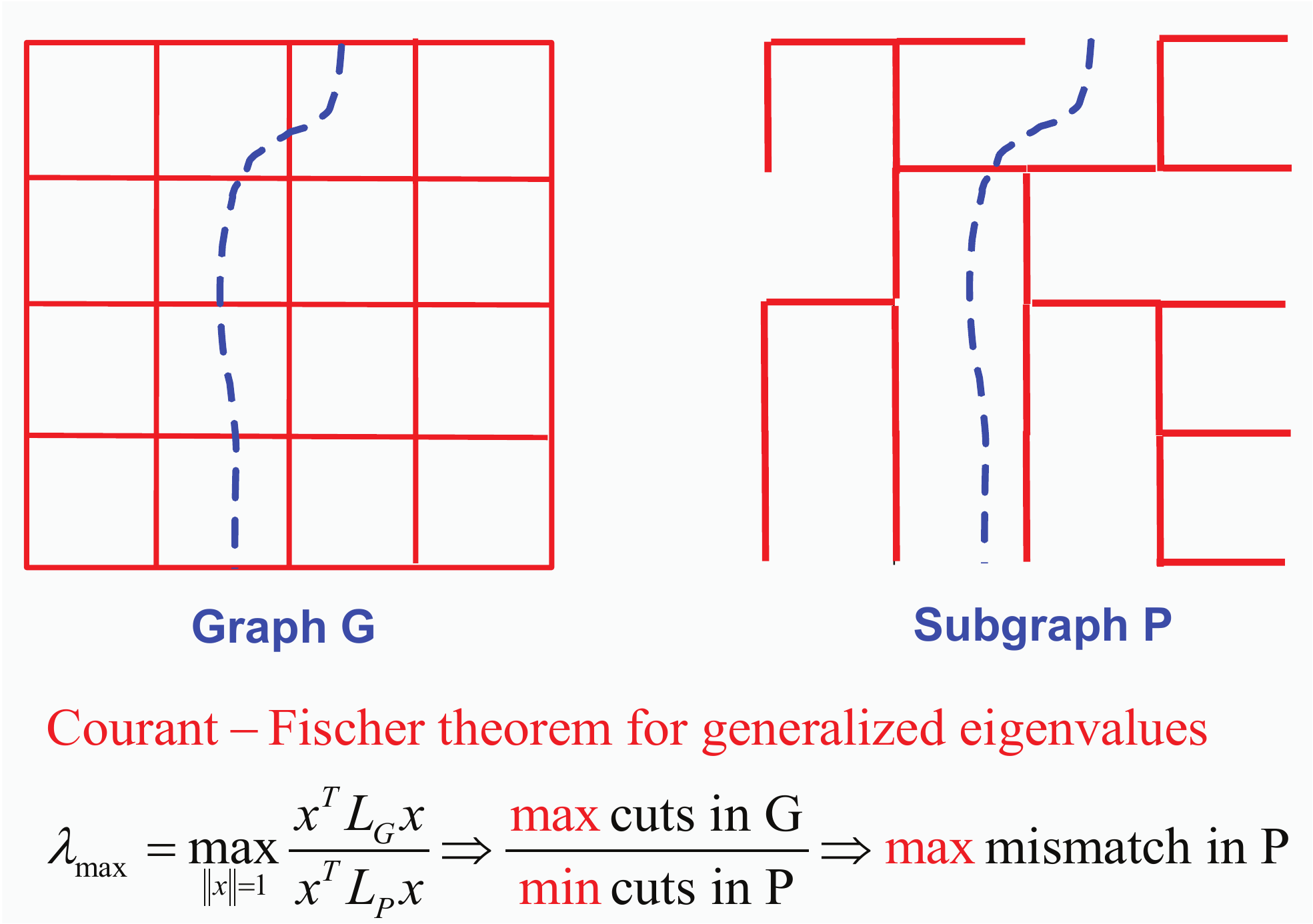, scale=0.4295} \caption{An alternative view based on Courant-Fischer theorem. \protect\label{fig:fischer}}
\end{figure}
 As a result, $\lambda_{max}=\lambda_{1}$ becomes the upper bound of the    mismatch in boundary (cut) size between $G$ and $P$, as shown in
 Fig. \ref{fig:fischer}.  
  \subsection{Problem Formulation}
 Once $Q$ or $\partial_G(Q)$ is found  using dominant generalized eigenvectors, we can recover a small number of  edges from $\partial_G (Q)$   to $P$ in order to dramatically  reduce the maximum mismatch ($\lambda_{1}$), and thus improve the spectral approximation   of $P$. To this end, we propose the following {problem formulation} for spectral graph sparsification:
 \begin{equation}\label{formula_spectral}
 \min_{\mathbf{L}_P}{\Bigg\{\max_{\mathbf{x}}\left(\frac{\mathbf{x}^\top \mathbf{L}_G \mathbf{x}}{\mathbf{x}^\top \mathbf{L}_P \mathbf{x}}\right)+\beta {{\|\mathbf{L}_P\|}}^{}_{1}\Bigg\}},
\end{equation}
  where $\mathbf{L}_P$ denotes Laplacian matrix of the subgraph $P$,  $\mathbf{x}^\top\mathbf{1}=0$ and $|\mathbf{x}|\neq 0$.
 (\ref{formula_spectral}) aims to minimize the largest generalized eigenvalue by adding the minimum amount of edges into the subgraph $P$, which can be solved iteratively by repeating the following two steps: 1) compute the generalized eigenvector corresponding to the largest (dominant) eigenvalue, and 2) identify  the most spectrally-critical off-tree edges that can mostly decrease the dominant eigenvalue(s) and add them into the subgraph $P$. Once the dominant eigenvalue is small enough (e.g., $\lambda=10$),   $P$ will be highly spectrally similar to the original graph $G$. 
 
However, computing the dominant eigenvalue and its eigenvector can sometimes be  too costly for large-scale graph Laplacian matrices, even when state-of-the-art   eigenvalue decomposition methods are adopted \cite{saad2011eigbook}. Additionally, there can still be too many eigenvalues  to be fixed in order to achieve a  desired spectral similarity level (relative condition number).

\subsection{Edge Embedding with Approximate Dominant Eigenvector}
To   more efficiently identify critical off-tree edges, approximate dominant generalized eigenvectors and eigenvalues of (\ref{formula_generalized_eigen}) can be exploited. To this end,  a generalized power iteration procedure that can be computed in nearly-linear time has been proposed in \cite{zhuo:dac16}. In the rest of this paper, we assume that a very small positive  diagonal element is added to a randomly selected row of the graph Laplacian matrix to convert the Laplacian matrix into a full-rank matrix.  We express an initial random vector $\mathbf{{h_0}}$ that is orthogonal to the all-one vector  using generalized eigenvectors $\mathbf{u_i}$ as follows:
\begin{equation}\label{formula_rand_vec}
\mathbf{{h_0} = \sum\limits_{i = 1}^n {{\alpha _i}{u_i}}},
\end{equation}
where $\mathbf{1^\top {h_0} }=0$.
Applying ${t}$-step power iterations to the   generalized eigenvalue problem, we have
\begin{equation}\label{formula_pwr_iter}
\mathbf{{h_t} = \left({L_P^{-1}}L_G\right)^t{h_0} = \sum\limits_{i = 1}^n {{\alpha _i}{\lambda^t _i}{u_i}}}.
\end{equation}
The  Laplacian quadratic form function $Q_{\mathbf{\delta L_P}}(\mathbf{h_t})=\mathbf{{h_t}^\top\delta L_P{h_t}}$ becomes:
\begin{equation}\label{formula_pwr_iter2}
\begin{array}{l}
Q_{\mathbf \delta L_P}(\mathbf{h_t})=\mathbf{{h^\top_t}\delta L_P{h_t}} = {\left( {\sum\limits_{i = 1}^n {{\alpha _i}{\lambda^t _i}\mathbf{{u_i}}} } \right)^\top}\delta \mathbf{L_P}\left( {\sum\limits_{j = 1}^n {{\alpha _j}{\lambda^t _j}\mathbf{{u_j}}} } \right)\\
 = \mathbf{\sum\limits_{i = 1}^n {{{\left( {{\alpha _i}{\lambda^t _i}} \right)}^2}{u^\top_i}} \delta L_P{u_i} + \sum\limits_{i=1}^n {\sum\limits_{j=1,j\ne i}^n { {{\alpha _i}{\alpha _j}{\lambda^t _i}{\lambda^t _j}} {u^\top_j} } \delta L_P{u_i}}}.
\end{array}
\end{equation}
Substituting (\ref{formula_deltaP}) into (\ref{formula_pwr_iter2}), $Q_{\mathbf{\delta L_P}}(\mathbf{h_t})$ becomes:
\begin{equation}\label{formula_pwr_iter5b}
\begin{split}
Q_{\mathbf \delta L_P}(\mathbf{h_t})&= \sum\limits_{(p,q)\in E\setminus E_s}^{} \omega_{p,q}\left(\sum\limits_{i = 1}^n {\alpha _i}{\lambda^t _i} \mathbf{{u^\top_i}  e_{p,q}}\right)^2\\
&=\sum\limits_{(p,q)\in E\setminus E_s}^{} \omega_{p,q} \overline{v}^2_{p,q},
\end{split}
\end{equation}
which indicates that an off-tree edge with greater $|\overline{v}_{p,q}|$ or Joule heat $\omega_{p,q} \overline{v}^2_{p,q}$  for $t>0$ will be a more spectrally-critical off-tree edge that can significantly influence  large  generalized eigenvalues. Expanding  (\ref{formula_pwr_iter5b}) leads to:
\begin{equation}\label{formula_pwr_iter5c}
\begin{array}{l}
Q_{\mathbf \delta L_P}(\mathbf{h_t}) = \sum\limits_{i = 1}^n {\alpha^2_i}{\lambda^{2t} _i} \sum\limits_{(p,q)\in E\setminus E_s}^{} \omega_{p,q}\mathbf{\left({u^\top_i}  e_{p,q}\right)^2+}\\
\sum\limits_{i=1}^n {\sum\limits_{j=1,j\ne i}^n}  {{\alpha _i}{\alpha _j}{\lambda^t _i}{\lambda^t _j}} \sum\limits_{(p,q)\in E\setminus E_s}^{} \omega_{p,q}\mathbf{{u_i}^\top e_{p,q}{u^\top_j}  e_{p,q}}.
\end{array}
\end{equation}
If we further define:
 \begin{equation}\label{formula_deltaPmax}
\delta \mathbf{L_{P, {max}}= L_G-L_P},
\end{equation}
 which can be considered as an extreme-case Laplacian matrix that includes all off-tree edges that  belong to the original graph $G$ but  not  the subgraph $P$, then the following equation holds:
  \begin{equation}\label{formula_pwr_iter5d}
\begin{array}{l}
\sum\limits_{i=1}^n {\sum\limits_{j=1,j\ne i}^n}  {{\alpha _i}{\alpha _j}{\lambda^t _i}{\lambda^t _j}} \sum\limits_{(p,q)\in E\setminus E_s}^{} w_{p,q}\mathbf{{u^\top_i}  e_{p,q}{u^\top_j}  e_{p,q}}=\\
\sum\limits_{i=1}^n {\sum\limits_{j=1,j\ne i}^n}  {{\alpha _i}{\alpha _j}{\lambda^t _i}{\lambda^t _j}} \mathbf{u^\top_i(L_G-L_P)u_j=0},
\end{array}
\end{equation}
which leads to the edge-based expansion of the  quadratic form for $\mathbf{\delta L_{P, {max}}}$ as follows:
  \begin{equation}\label{formula_pwr_iter5new}
  \begin{array}{l}
Q\mathbf{_{\delta L_{P, {max}}}(h_t)={h^\top_t}\delta L_{P, {max}}{h_t} = \sum\limits_{i = 1}^n {{{\left( {{\alpha _i}{\lambda^t _i}} \right)}^2(\lambda_i-1)}}}\\=\sum\limits_{(p,q)\in E\setminus E_s}^{} w_{p,q}\sum\limits_{i = 1}^n{\alpha^2_i}{\lambda^{2t} _i}\mathbf{\left({u^\top_i} e_{p,q}\right)^2}.
\end{array}
\end{equation}

 It is obvious that (\ref{formula_pwr_iter5b}), or (\ref{formula_pwr_iter5new}) will allow ranking each off-tree edge according to its spectral-criticality level. (\ref{formula_pwr_iter5new})  also indicates that adding all off-tree edges with nonzero Joule heat values back to $P$ will immediately bring all generalized eigenvalues to $1$. It should be noted that the required number  of generalized power iterations  can be rather small (e.g. $t=2$)  in practice  to observe good result.


%

\subsection{Spectrally-unique Off-tree Edges }\label{sec_rank1_Pmat}
Define a spectrally-unique off-tree edge ($p_i,q_i$)  to be the off-tree edge  that can  completely and only fix one  large (dominant) generalized eigenvalue $\lambda_i$, though each edge  will usually influence more than one eigenvalues and eigenvectors   according to (\ref{formula_pwr_iter5new}). Then the following truncated expansion of the Laplacian quadratic form can be obtained  when considering the top $k$  most dominant yet spectrally-unique  off-tree edges for fixing the   top $k$ largest eigenvalues:
 \begin{equation}\label{formula_eigfix}
 \begin{array}{l}
{Q_{\mathbf\delta L_{P, {max}}}(\mathbf{ h_t}) \approx \sum\limits_{i = 1}^{k} w_{p_i,q_i}{\alpha^2_i}{\lambda^{2t}_i}\left({u^\top_i}   e_{p_i,q_i} \right)^2}\\
~~~~~~~~~~~~~~~~=\sum\limits_{i = 1}^k {{{ {{\alpha^2 _i}{\lambda^{2t} _i}} }(\lambda_i-1)}},
\end{array}
\end{equation}
 where we should be able to find $\gamma_i\ne 0$ such that:
\begin{equation}\label{formula_p-orth2}
\mathbf{e_{p_i,q_i}=\gamma_i L_P u_i},
\end{equation}
while the following will also be satisfied:
 \begin{equation}\label{formula_p-orth3}
\mathbf{u_j^\top e_{p_i,q_i}}= \left\{ \begin{array}{l}
\gamma_i,i = j,\\
0,i \ne j.
\end{array} \right.\end{equation}
 Obviously, for random coefficients $\alpha_i$, the following must be satisfied for a spectrally-unique off-tree edge:
 \begin{equation}\label{formula_eigfix3}
 w_{p_i,q_i}\gamma^2_i= { {\lambda_i-1} }.
\end{equation}
Consequently, if there is an off-tree edge $(p_i,q_i)$ with weight $w_{p_i,q_i}$ that satisfies:
\begin{equation}\label{formula_p-orth4}
\mathbf{e_{p_i}-e_{q_i}}=\pm\sqrt{\frac{\lambda_i-1}{w_{p_i,q_i}}} \mathbf{L_P u_i},
\end{equation}
adding this off-tree edge back to the spanning tree will completely fix the corresponding eigenvalue $\lambda_i$.
Then the effective resistance of edge $(p_i, q_i)$ in $P$ becomes:
  \begin{equation}\label{formula_Reff}
R^{\mathbf{eff}}_{e_i}=\mathbf{e}^\top_{p_i,q_i}\mathbf{L^+_P e}_{p_i,q_i}=\gamma^2_i \mathbf{u^\top_i L_P u_i}=\gamma^2_i,
\end{equation}
which immediately leads to:
  \begin{equation}\label{formula_eigfix2}
Q\mathbf{_{\delta L_{P, {max}}}(h_t)} \approx  \sum\limits_{i = 1}^{k}  {\alpha^2_i}{\lambda^{2t}_i} w_{p_i,q_i} R^{\mathbf{eff}}_{e_i}\approx\sum\limits_{i = 1}^k {{{ {{\alpha^2 _i}{\lambda^{2t+1} _i}} }}}.
\end{equation}

Since the stretch of off-tree edge $(p_i, q_i)$ is computed by ${{\rm{s}}{{\rm{t}}_{P}}\left( p_i, q_i \right)}=w_{p_i,q_i} R^{\mathbf{eff}}_{e_i}$, (\ref{formula_eigfix2}) also indicates that ${{\rm{s}}{{\rm{t}}_{P}}\left( e_i \right)}\approx\lambda_i$ holds for spectrally-unique off-tree edges. Consequently, the key off-tree edges identified by (\ref{formula_pwr_iter5new}) or (\ref{formula_eigfix2}) will have the largest stretch values  and therefore most significantly impact the largest eigenvalues of $\mathbf{L_P^{+}L_G}$.  (\ref{formula_eigfix2}) also can be considered as a randomized version of  $\mathbf{Trace({L_P^{+}}L_G)}$ that is further scaled up by a factor of $\lambda^{2t}_{i}$.

\subsection{Rank-One Update with A Spectrally-Unique   Edge }\label{sec_rank1_update}
 The updated generalized eigenvalue $\lambda'_i$ after adding one spectrally-unique off-tree edge $(p_i,q_i)$ with weight $w_{p_i,q_i}$ back to the spanning tree for fixing  eigenvalue $\lambda_i$ can be derived based on the Sherman-Morrison Formula and Matrix Determinant Lemma. Define matrix $A_P$ to be:
\begin{equation}\label{formula_rank_one_update1}
\mathbf{A_P=L_G^{\frac{1}{2}}L_P^{-1}L_G^{\frac{1}{2}}},
\end{equation}
which has the same set of eigenvalues of matrix $\mathbf{L_P^{-1}L_G}$.
After adding the off-tree edge $(p_i,q_i)$, the updated $\mathbf{A_P}$ is denoted by $\mathbf{A'_P}$ that is expressed as:
\begin{equation}\label{formula_rank_one_update}
\begin{array}{l}
\mathbf{A'_P=L_G^{\frac{1}{2}}}\left(\mathbf{L_P}+w_{p_i,q_i} \mathbf{e_{p_i,q_i} e_{p_i,q_i}}^\top\right)^{-1}\mathbf{L_G^{\frac{1}{2}}}\\= \mathbf{A_P}-\frac{w_{p_i,q_i}\mathbf{L_G^{\frac{1}{2}}L_P^{-1}e_{p_i,q_i}e_{p_i,q_i}^\top L_P^{-1}L_G^{\frac{1}{2}}}}{1+w_{p_i,q_i}\mathbf{e_{p_i,q_i}^\top L_P^{-1}e_{p_i,q_i}}}=\mathbf{A_P-v_Pv_P^\top},
\end{array}
\end{equation}
where $\mathbf{v_P}$ is defined as:
\begin{equation}\label{formula_rank_one_update2}
\mathbf{v_P}=\frac{\sqrt{w_{p_i,q_i}}\mathbf{L_G^{\frac{1}{2}}L_P^{-1}e_{p_i,q_i}}}{\sqrt{1+w_{p_i,q_i}\mathbf{e_{p_i,q_i}^\top L_P^{-1}e_{p_i,q_i}}}}=\frac{\gamma_i\sqrt{w_{p_i,q_i}}\mathbf{L_G^{\frac{1}{2}}u_i}}{\sqrt{1+w_{p_i,q_i}\gamma_i^2}}.
\end{equation}
The characteristic polynomial of $\mathbf{A'_P=A_P-v_Pv_P^\top }$ can be computed as follows:
\begin{equation}\label{formula_determinant}
\begin{array}{l}
p_{\mathbf{A'_P}}(x)=p_{\mathbf{A_P - v_P v_P^\top }}(x)=\det(\mathbf{xI-A_P + v_P v_P^\top })\\
=\det(\mathbf{xI-A_P})\det\left(\mathbf{I+(xI-A_P)^{-1}v_P v_P^\top }\right).
\end{array}
\end{equation}
According to Matrix Determinant Lemma, we have:
\begin{equation}\label{formula_determinant2}
p_{\mathbf{A'_P}}(x)=p_{\mathbf{A_P}}(x)\left(1+\mathbf{v_P^\top (xI-A_P)^{-1} v_P}\right).
\end{equation}
Denoting $\mathbf{z_i}$ for $i=1,...,n$ the unit-length orthonormalized eigenvectors of matrix $\mathbf{A_P}$ that correspond  to eigenvalues $\lambda_i$ respectively,  we have:
\begin{equation}\label{formula_determinant3}
(\mathbf{xI-A_P})^{-1}=\left( \sum\limits_{i = 1}^{n}(x-\lambda_i)\mathbf{z_i z_i^\top }\right)^{-1}= \sum\limits_{i = 1}^{n}\frac{\mathbf{z_i z_i^\top }}{x-\lambda_i},
\end{equation}
which leads to:
 \begin{equation}\label{formula_determinant4}
p_{\mathbf{A'_P}}(x)=p_{\mathbf{A_P}}(x)\left(1+\sum\limits_{i = 1}^{n}\frac{(\mathbf{v_P^\top z_i})^2}{x-\lambda_i}\right).
\end{equation}
It has been shown in \cite{batson2012twice}  that (\ref{formula_determinant4}) indicates: 1) the latest eigenvalues after rank-one update will always be reduced if  $(\mathbf{v_P^\top  z_i})^2>0$; 2) the greater value of $(\mathbf{v_P^\top  z_i})^2$ will result in greater reduction in $\lambda_i$. Therefore, if an off-tree edge satisfies  $(\mathbf{v_P^\top  z_i})^2>>0$ for multiple eigenvectors $\mathbf{z_i}$, adding this edge back to the spanning tree subgraph will substantially reduce multiple eigenvalues at the same time; on the other hand, a spectrally-unique off-tree edge will substantially reduce only one eigenvalue. It can be shown that:
\begin{equation}\label{formula_rank_one_eig}
\mathbf{L_G^{\frac{1}{2}}L_P^{-1}L_G^{\frac{1}{2}}\left(\frac{L_G^{\frac{1}{2}}u_i}{\|L_G^{\frac{1}{2}}u_i\|}\right)=\lambda_i \left(\frac{L_G^{\frac{1}{2}}u_i}{\|L_G^{\frac{1}{2}}u_i\|}\right)=\lambda_i z_i},
\end{equation}
which indicates that $\mathbf{v_P}$ is an eigenvector of  matrix $\mathbf{A_P}$ corresponding to the eigenvalue $\lambda_i$.
Consequently, the characteristic polynomial of $\mathbf{A'_P=A_P-v_P v_P^\top }$ can be further simplified into the following form:
\begin{equation}\label{formula_rank_one_update3}
p_{\mathbf{A'_P}}(x)=p_{\mathbf{A_P}}(x)\left(1+\frac{(\mathbf{v_P^\top z_i})^2}{x-\lambda_i}\right)=p_{\mathbf{A_P}}(x)\left(1+\frac{\mathbf{v_P^\top v_P}}{x-\lambda_i}\right).
\end{equation}
 Therefore, the updated eigenvalue $\lambda'_i$ after adding the off-tree edge can be computed by solving $p_{A'_P}(x)=0$, which leads to:
\begin{equation}\label{formula_rank_one_update4}
\lambda'_i=\lambda_i-\mathbf{v_P^\top v_P}=\frac{\lambda_i}{1+ w_{p_i,q_i}\gamma_i^2}.
\end{equation}
For achieving the desired $\lambda'_i$ after adding the off-tree edge, the edge weight should be set as:
\begin{equation}\label{formula_rank_one_update5}
w_{p_i,q_i}=\frac{\lambda_i-\lambda'_i}{\lambda'_i\gamma_i^2}.
\end{equation}
It can be shown that when the desired $\lambda'_i=1$, we have $w_{p_i,q_i}=\frac{\lambda_i-1}{\gamma_i^2}$, which is equivalent to (\ref{formula_eigfix3}).
As a result, considering the nearly-worst case eigenvalue distribution $\lambda _{i}\le{{\rm{s}}{{\rm{t}}_{P}}\left( G \right)} /i$ shown in Fig. \ref{fig:eigworst}, a $\sigma$-similar spectral sparsifier with $n-1+O(\frac{m \log n \log \log n}{\sigma^2})$ edges can be obtained in $O(m)$ time using the proposed method when  an initial low-stretch spanning tree is given.

 \vspace{-0.1951500cm}
\subsection{Algorithm Flow and Complexity}\label{sec_iter_sparse}
The detailed algorithm flow of the proposed  spectral graph sparsification approach has been summarized as follows:
\begin{enumerate}
  \item Extract a spanning tree subgraph  (e.g. a scaled low-stretch spanning tree \cite{elkin2008lst,abraham2012}) from the original graph;
  \item Perform $t$-step generalized power iterations to compute $h_{t}$ with an initial random vector;
  \item   Compute the spectral criticality of each edge based on the  Laplacian quadratic form of $\delta L_{P, {max}}$ using (\ref{formula_pwr_iter5new});
   \item  Rank each  edge  using its spectral criticality levels;
  \item  Add a small portion of dissimilar off-tree edges  back to the   spanning tree  to form the   ultra-sparse spectral graph sparsifier.
  \end{enumerate}

 It should  be noted that the proposed spectral graph sparsification approach allows  ranking all off-tree edges according to their spectral criticality levels in a very efficient and effective way. Compared to the state-of-the-art sampling-based approaches that rely on effective resistance calculations, the proposed method can achieve the very similar goal of ranking spectrally critical off-tree  edges.

 The complexity of the proposed spectral perturbation based approach can be analyzed by considering two key steps: (a) spanning tree construction based on the original graph, and (b) ultra-sparsifier construction based on the spanning tree. Recent research has shown that low-stretch  spanning trees in (a) can be constructed in nearly-linear time \cite{elkin2008lst, abraham2012}. For instance, the petal-decomposition algorithm requires $O(m\log n \log \log n)$ time to generate a low-stretch spanning tree with a total stretch of $O(m \log n \log \log n)$\cite{abraham2012}; the  $t$-step generalized power iterations in (b) can be achieved in linear time for a fixed $t$ since  factorization of a tree-graph Laplacian matrix can be accomplished within linear $O(m)$ time. Consequently, the overall complexity of the proposed spectral perturbation based sparsification algorithm is almost linear.
 
 \section{Similarity-Aware Spectral  Sparsification by Edge Filtering}\label{sec:similarity}
 Although  (\ref{formula_pwr_iter5new}) and  (\ref{formula_eigfix2})  provide  a  spectral ranking for each off-tree edge, it is not clear how many off-tree  edges should be recovered to the spanning tree for achieving a desired spectral similarity level. To this end, we introduce  a simple yet effective spectral off-tree edge filtering scheme motivated by recent graph signal processing techniques \cite{shuman2013emerging}. 
 \subsection{Spectral  Sparsification: A Low-Pass Filter on Graphs}
\begin{figure}
\centering \epsfig{file=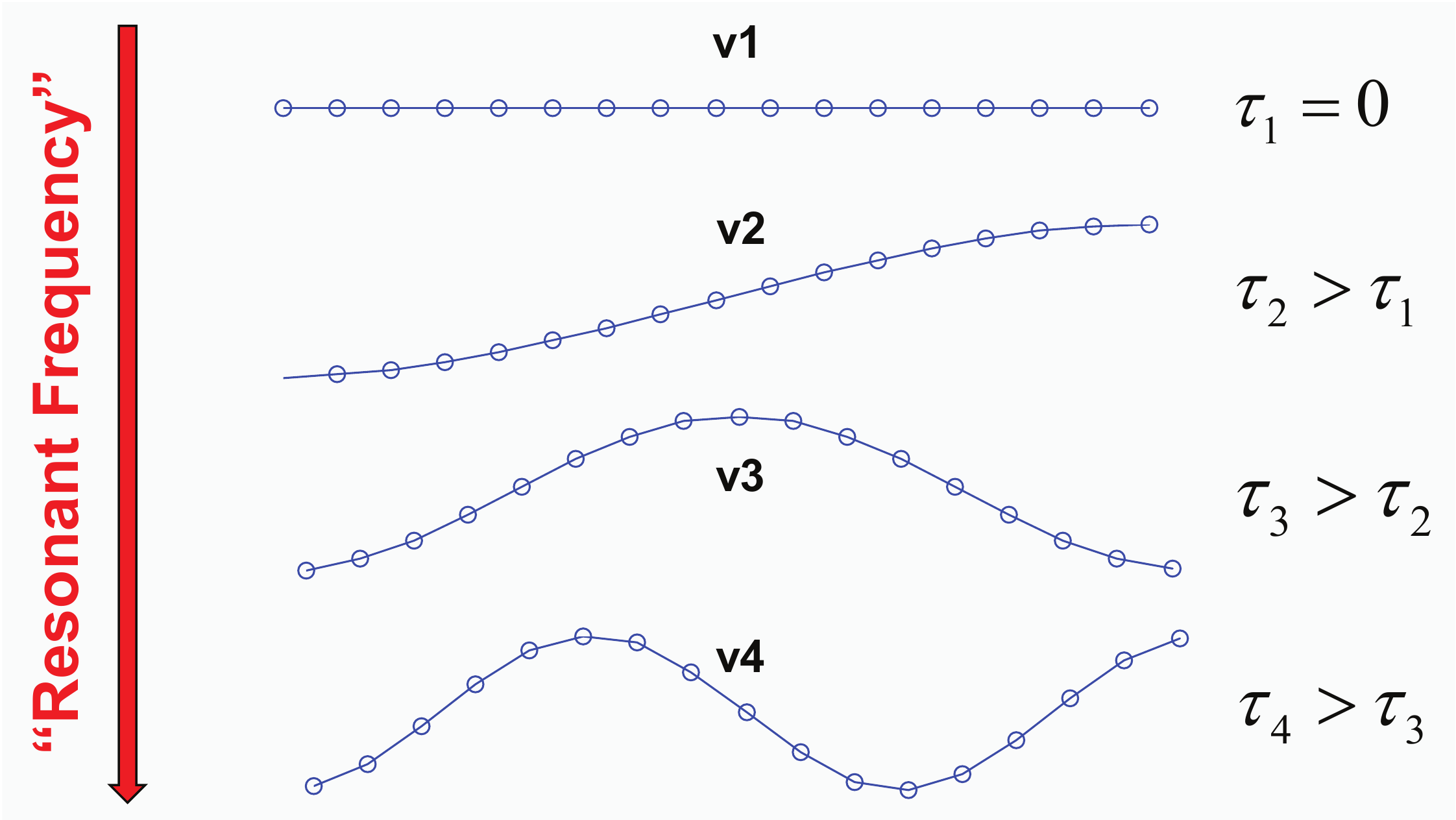, scale=0.39} \caption{The  eigenvectors ($v_i$) of increasing eigenvalues ($\tau_i$) for a path graph. \protect\label{fig:linegraph}}
\end{figure}

To more efficiently analyze signals on general undirected graphs,   graph signal processing techniques have been extensively studied recently \cite{shuman2013emerging}. There is a clear analogy between traditional signal processing based on classical Fourier analysis and graph signal processing: 1) the signals at different time points in classical Fourier analysis correspond to the signals at different  nodes in an undirected graph; 2) the more slowly oscillating functions   in time domain correspond to the graph Laplacian eigenvectors associated with lower eigenvalues and more slowly varying (smoother) components across the graph. 
For example, the first  few nontrivial eigenvectors associated with the smallest non-zero eigenvalues of a path graph Laplacian have been illustrated in Fig. \ref{fig:linegraph}, where the increasing eigenvalues correspond to increasing oscillation frequencies in the path graph. 
A comprehensive review of fundamental signal processing operations, such as filtering, translation, modulation, dilation, and down-sampling to the graph setting has been provided in \cite{shuman2013emerging}.

Spectral sparsification  aims to  maintain a simplest subgraph  sufficient for preserving the slowly-varying  or ``low-frequency" signals on graphs, which therefore can be regarded as a ``low-pass" graph filter. In other words, such spectrally sparsified graphs will be able to preserve the eigenvectors associated with low eigenvalues more accurately than high eigenvalues, and thus will retain ``low-frequency" graph signals sufficiently well, but  not so well for highly-oscillating (signal) components due to the  missing edges. 

In practice, preserving the spectral (structural) properties  of the original graph within the spectral sparsifier is  key to design of  many fast  numerical and graph-related  algorithms \cite{spielman2011graph,miller:2010focs, christiano2011flow, spielman2014sdd}. For example, when using spectral sparsifier as a preconditioner in preconditioned conjugate gradient (PCG) iterations, the convergence rate only depends on the  spectral similarity (or relative condition number) for achieving a desired accuracy level, while in spectral graph partitioning and data clustering tasks only the first few eigenvectors associated with the smallest nontrivial eigenvalues of graph Laplacian are needed \cite{spielmat1996spectral,peng2015partitioning}. 
 
 \subsection{Off-Tree Edge Filtering with Joule Heat}
  To only recover the off-tree edges that are most critical for achieving the desired spectral similarity level, we propose the following scheme for truncating spectrally-unique off-tree edges based on each edge's Joule heat. For a spanning-tree preconditioner, since there will be at most $k$ generalized eigenvalues that are greater than ${{\rm{s}}{{\rm{t}}_{P}}\left( G \right)} /k$, the following simple yet nearly worst-case generalized eigenvalue distribution can be assumed:
  
\begin{equation}\label{formula_eig_dist}
\lambda _{i}= \frac{2\lambda_{max}}{i+1}=\frac{{\rm{s}}{{\rm{t}}_{P}}\left( G \right)}{i+1},i\geq 1.
\end{equation}
To  most economically select  the top-$k$ spectrally-unique off-tree edges that will dominantly impact the top-$k$ largest generalized eigenvalues, the  following sum of quadratic forms  (Joule heat levels) can be computed based on (\ref{formula_eigfix2}) by performing $t$-step generalized power iterations with $r$ multiple random vectors  $\mathbf{h_{t,{1}},...,h_{t,{r}}}$:
\begin{equation}\label{formula_pwr_iter6}
Q\mathbf{_{\delta L_{P, {max}}}(h_{t,{1}},...,h_{t,{r}})}   \approx  \sum\limits_{j = 1}^r \sum\limits_{i = 1}^k {{{(\alpha_{i,j})^2\left( \frac{2\lambda_{max}}{i+1} \right)}^{2t+1}}}.
\end{equation}
The goal is to select top $k$  spectrally-unique  off-tree edges  for fixing the top  $k$ largest generalized eigenvalues such that  the resulting upper bound of the relative condition number will become ${\sigma}^2=\frac{\tilde{\lambda}_{max}}{\tilde{\lambda}_{min}}$, where $\tilde{\lambda}_{max}$ and $\tilde{\lambda}_{min}$ denote the largest and smallest eigenvalues of $\mathbf{L_P^+L_G}$ after adding top-k spectrally-unique off-tree edges. Then we have:
\begin{equation}\label{formula_eig_k}
k=2\lambda_{max}/\tilde{\lambda}_{max}-1.
\end{equation}
When using multiple random vectors for computing (\ref{formula_pwr_iter6}), it is expected that ${{\sum\limits_{j = 1}^r \alpha _{k,j}^2}}\approx {{\sum\limits_{j = 1}^r \alpha _{1,j}^2}}$, which allows us to define the normalized edge Joule heat $\theta _k$ for the $k$-th spectrally-unique  off-tree edge  through the following simplifications:
  \begin{equation}\label{formula_heatratio}
 {\theta _k} = \frac{{{heat_{{\lambda _k}}}}}{{{heat_{{\lambda _1}}}}} = {\left( {\frac{{{\sum\limits_{j = 1}^r \alpha _{k,j}^2}}}{{{\sum\limits_{j = 1}^r \alpha^2 _{1,j}}}}} \right)}{\left( \frac{\tilde{\lambda}_{max} }{ \lambda_{max}} \right)^{2t+1}}\approx\left( \frac{{\sigma}^2  \tilde{\lambda}_{min} }{ \lambda_{max}} \right)^{2t+1}.
\end{equation}
The key idea of the proposed similarity-aware spectral sparsification is to leverage the normalized Joule heat (\ref{formula_heatratio}) as a  threshold for filtering off-tree edges: only the off-tree edges with normalized Joule heat values greater than $\theta _k$ will be selected for inclusion into the spanning tree for achieving the desired spectral similarity  ($\sigma$) level.
Although the above  scheme is derived for filtering spectrally-unique off-tree edges,   general off-tree edges also can be filtered using similar strategies. Since adding the off-tree edges with largest Joule heat to the subgraph will mainly impact the largest generalized eigenvalues but not the smallest ones, we will  assume $\tilde{\lambda}_{min}\approx {\lambda}_{min}$, and  use the following edge truncation scheme for filtering general off-tree edges: the off-tree edge $(p,q)$ will be included into the sparsifier if its normalized Joule heat value is greater than the  threshold determined by:
\begin{equation}\label{formula_heatratio_final}
\theta _{(p,q)}=\frac{heat_{(p,q)} }{heat_{max}} \ge \theta_{{\sigma}} \approx \left( \frac{{\sigma}^2 {\lambda}_{min}}{ \lambda_{max}} \right)^{2t+1},
\end{equation}
where $\theta_{{\sigma}}$ denotes the threshold for achieving the $\sigma-$spectral similarity in the sparsifier, and $heat_{max}$ denotes the maximum  Joule heat  of all off-tree edges computed by (\ref{formula_pwr_iter5new}) with multiple initial random vectors.  

\subsection{Estimation of Extreme Eigenvalues}\label{sec_extreme-eigvalue}
To achieve the above spectral off-tree edge filtering scheme, we need to compute $\theta_{{\sigma}} $ in (\ref{formula_heatratio_final}) that further requires to estimate the extreme    eigenvalues  $\lambda_{max}$ and $\lambda_{min}$ of $\mathbf{L_P^+L_G}$. In this work, we propose the following efficient methods for computing these extreme generalized eigenvalues.
\subsubsection{Estimating  $\lambda_{max}$ via Power Iterations } 
 Since generalized power iterations  converge at a geometric rate determined by the separation of the two largest generalized eigenvalues $\lambda_{max}=\lambda_1>\lambda_2$ , the error of the estimated  eigenvalue  will decrease quickly when $|\lambda_2/\lambda_1|$ is small. It has been shown that the largest eigenvalues of $\mathbf{L_P^{+}L_G}$ are well separated from each other \cite{spielman2009note}, which thus leads to very fast convergence of generalized power iterations for estimating  $\lambda_{1}$. To achieve scalable performance of power iterations, we can adopt  recent graph-theoretic algebraic multigrid (AMG) methods for  solving the sparsified Laplacian matrix $\mathbf{L_P}$  \cite{livne2012lean, zhiqiang:iccad17}. 
 
\subsubsection{Estimating  $\lambda_{min}$ via Node Coloring} 
Since the smallest eigenvalues of $\mathbf{L_P^{+}L_G}$ are crowded together \cite{spielman2009note}, using (shifted) inverse power iterations may not be efficient due to the extremely slow convergence rate. To the extent of our knowledge, none of existing eigenvalue decomposition methods can  efficiently compute $\lambda_{min}$.  

This work exploits the following Courant-Fischer theorem for generalized eigenvalue problems: 
\begin{equation}\label{formula_courant-fischer}
\lambda_{min}=\min_{|\mathbf{x}| \neq 0}\frac{\mathbf{x^\top  L_G {x}}}{\mathbf{x^\top  L_P {x}}},
\end{equation}
where $\mathbf{x}$ is also required to be orthogonal to the all-one vector. (\ref{formula_courant-fischer}) indicates that if we can find a vector $\mathbf{x}$ that minimizes the ratio between the quadratic forms of the original and sparsified Laplacians, $\lambda_{min}$ can be subsequently computed. By restricting the values in $\mathbf{x}$ to be only  $1$ or $0$, which can be considered as assigning one of the two colors to each node in graphs $G$ and $P$,  the following   simplifications can be made:
\begin{equation}\label{formula_courant-fischer-2}
\lambda_{min}\le\mathop{\min_{\mathbf{|x|} \neq 0}}_{x(i)\in \left\{0,1 \right\}}\mathbf{\frac{x^\top  L_G x}{x^\top  L_P x}}=\mathop{\min_{\mathbf{|x|} \neq 0}}_{x(i)\in \left\{0,1 \right\}}{\frac{\sum\limits_{x(p)\neq x(q), (p,q)\in E} w_{pq}}{\sum\limits_{x(p)\neq x(q), (p,q)\in E_s} w_{pq}}},
\end{equation}
which will always allow estimating an upper bound for $\lambda_{min}$.  To this end, we first initialize all nodes with $0$ value and subsequently try to find a node $p$ such that  the ratio between quadratic forms can be minimized:
\begin{equation}\label{formula_courant-fischer-3}
\lambda_{min}\le \min_{p\in V}{\frac{ L_G(p,p)}{L_P(p,p)}}.
\end{equation}
The above procedure for estimating $\lambda_{min}$ only requires finding the node with the smallest node degree ratio and thus can be easily implemented and efficiently performed for even very large graph problems.  Our  results for real-world  graphs show that the proposed method is highly efficient and can very well estimate the smallest generalized eigenvalues when compared with existing generalized eigenvalue  methods \cite{saad2011eigbook}.

\subsection{Iterative Sparsification for Ill-Conditioned Problems}\label{sec_iter_sparse2}
To achieve more effective edge filtering for  similarity-aware spectral graph sparsification, we propose to iteratively recover off-tree edges to  the sparsifier through an incremental graph densification procedure. Each densification iteration adds a small portion of ``filtered" off-tree edges to the latest spectral sparsifier, while the  spectral similarity is estimated to determine if more off-tree edges are needed. The  $i$-th graph densification iteration includes the following steps:
\begin{enumerate}
   \item Update the subgraph Laplacian matrix $\mathbf{L_P}$ as well as its    solver by leveraging recent graph-theoretic algebraic multigrid methods \cite{livne2012lean,zhiqiang:iccad17};
   \item Estimate the spectral similarity  by computing $\lambda_{max}$ and $\lambda_{min}$ using the methods described in Section \ref{sec_extreme-eigvalue};
  \item If the spectral similarity is not satisfactory, continue with the following steps; otherwise, terminate the subgraph densification procedure.
 \item Perform $t$-step generalized power iterations with  $O(\log |V|)$ random vectors to compute the sum of Laplacian quadratic forms  (\ref{formula_pwr_iter6});
  \item Rank and filter each off-tree edge  according to its normalized Joule heat value using  the threshold  $\theta_{{\sigma}} $ in (\ref{formula_heatratio_final});
  \item  Check the similarity of each selected off-tree edge and only add dissimilar edges to the latest sparsifier.
\end{enumerate}
\section{Experimental results}\label{result_sec}
The proposed spectral perturbation based spectral graph sparsification method (GRASS) has been implemented in $C$++ and available for download \footnote{https://sites.google.com/mtu.edu/zhuofeng-graphspar}. The proposed spectral graph sparsification algorithm allows   developing nearly-linear time algorithms for tackling SDD or SDD-like sparse matrix problems, spectral graph partitioning problems, as well as graph-based regression problems \cite{zhu2003semi, miller:2010focs, spielman2014sdd,xueqian:dac12,lengfei:iccad13}. In this paper,  a sparse SDD matrix algorithm has been implemented and compared to  the state-of-the-art sparse matrix solver, Cholmod \cite{cholmod}. Test cases demonstrated in this paper cover a great variety of sparse SDD matrix problems obtained from realistic VLSI power grid problems \cite{ibmpg,thupg}, and the sparse matrix collection from the University of Florida that includes integrated circuit simulation problems, three-dimensional thermal analysis problems, finite-element analysis, etc \cite{davis2011matrix}. Additionally, a spectral graph partitioning engine is also implemented, which has been dramatically accelerated by  taking advantage of the proposed spectral sparsification approach. All experiments are performed using a single CPU core of a computing platform running 64-bit RHEW 6.0 with a $2.67$GHz 12-core CPU.

\subsection{A Scalable Iterative Solver For Power Grid Analysis}
The spectral sparsifier obtained by the proposed algorithm (without  weight re-scaling for the spanning tree and off-tree edges) is leveraged as a preconditioner in a PCG solver. The preconditioner is factorized by the same  Cholmod solver \cite{cholmod}. The right-hand-side (RHS) input vector $b$ is generated randomly and the solver is set to converge to  an accuracy level $||A x-b||<10^{-3}||b||$  for all test cases.  ``$|V|$" denotes the number of nodes, ``NNZ" denotes the number of nonzero elements in the original matrix, `` $T_{D}$" (`` $T_{I}$")  denotes  the total solution time including both the  matrix factorization and resolving steps of the direct (iterative) solver, `` $M_{D}$" (`` $M_{I}$")  denotes  the memory cost for sparse matrix factorizations, ``$N_I$" denotes the number of iterations for the PCG solver to converge to the required accuracy level,    $\frac{\lambda_1}{\lambda_{1,fin}}$ calculates the reduction rate of the largest eigenvalue using the proposed spectral sparsification approach when compared  to the initial spanning tree preconditioner, and $\frac{N^{ST}_I}{N_I}$ is the ratio of required iteration numbers using the initial spanning-tree preconditioners and the new subgraph preconditioners.

\begin{table}
\begin{center}
\scriptsize \addtolength{\tabcolsep}{-2.5pt} \centering
\caption{Results of sparse SDD matrix solver  (spectral  sparsification  with $5\%$ to $10\% |V|$ extra off-tree edges) for test cases in \cite{ibmpg}.}
\begin{tabular}{|c|c|c|c|c|c|c|c|}
 \hline    CKTs & $|V|$ & NNZ &  $T_{D}$ ($M_{D}$) &  $T_{I}$ ($M_{I}$) & $N_I$ & $\frac{\lambda_1}{\lambda_{1,fin}}$ & $\frac{N^{ST}_I}{N_I}$ \\
 \hline ibmpg3 & 0.9E6 &   3.7E6 & 15.0s (0.8G) & 1.2s (0.2G) &  13 & 37X & 6X \\
 \hline ibmpg4 & 1.0E6 &   4.1E6 & 18.3s (1.0G) & 1.3s (0.2G) & 12 & 18X& 4X \\
 \hline ibmpg5 & 1.1E6 &   4.3E6 & 12.7s (0.6G) & 1.3s (0.2G) & 12 & 2,826X& 50X \\
 \hline ibmpg6 & 1.7E6 & 6.6E6 & 18.3s (0.9G)& 2.5s (0.3G) & 13 & 173X & 13X\\
 \hline ibmpg7 & 1.5E6 & 6.2E6 & 27.2s (1.3G)& 2.3s (0.3G) & 13 & 177X & 13X\\
 \hline ibmpg8 & 1.5E6 & 6.2E6 & 18.7s (1.3G) & 2.3s (0.3G) & 13 & 120X & 11X\\
  \hline
\end{tabular}\label{table:ibmpg}
\end{center}
\end{table}

\begin{table}
\begin{center}
\scriptsize \addtolength{\tabcolsep}{-2.5pt} \centering
\caption{Results of sparse SDD matrix solver (spectral sparsification  with $1\%$ to $2\% |V|$ extra off-tree edges) for test cases in \cite{thupg}.}
\begin{tabular}{|c|c|c|c|c|c|c|c|}
 \hline    CKTs & $|V|$ & NNZ &  $T_{D}$ ($M_{D}$) &  $T_{I}$ ($M_{I}$) & $N_I$ & $\frac{\lambda_1}{\lambda_{1,fin}}$& $\frac{N^{ST}_I}{N_I}$\\
 \hline thupg1 & 5.0E6 &  2.1E7 & 75s (4.0G) & 10s  (0.8G) &  27 & 34,047X & 185X \\
 \hline thupg2 & 8.9E6 &   3.9E7 & 158s (7.6G) & 21s (1.5G) & 32 & 39,426X  & 199X\\
 \hline thupg3 & 1.2E7 &   5.1E7 & 250s (10.0G) & 25s (1.9G) & 32 & 101,052X  & 318X \\
 \hline thupg4 & 1.5E7 & 6.6E7 & N/A            & 36s (2.5G) & 32 & 97,550X  & 312X\\
 \hline thupg5 & 1.9E7 & 8.5E7 & N/A            & 47s (3.1G) & 33 & 136,678X  & 370X\\
 \hline thupg6 & 2.4E7 & 1.1E8 & N/A            & 62s (3.8G) & 34 & 108,898X  & 330X\\
 \hline thupg7 & 2.8E7 & 1.2E8 & N/A            & 70s (4.6G) & 34 & 87,463X  & 296X\\
 \hline thupg8 & 4.0E7 & 1.8E8 & N/A            & 110s (6.5G) & 34 & 368,898X  & 607X\\
  \hline
\end{tabular}\label{table:thupg}
\end{center}
\end{table}

\begin{table}
\begin{center}
\scriptsize \addtolength{\tabcolsep}{-2.5pt} \centering
\caption{Results of sparse SDD matrix solver (spectral  sparsification  with $5\%$ to $10\% |V|$ extra off-tree edges) for test cases in \cite{davis2011matrix}.}
\begin{tabular}{|c|c|c|c|c|c|c|c|}
 \hline   Test Cases & $|V|$ & NNZ &  $T_{D}$ ($M_{D}$) &  $T_{I}$ ($M_{I}$) & $N_I$ & $\frac{\lambda_1}{\lambda_{1,fin}}$ & $\frac{N^{ST}_I}{N_I}$\\
 \hline G3\_circuit & 1.6E6 &  7.7E6 & 45.1s (2.2G) & 5.2s  (0.3G) &  37 & 45,897X& 214X \\
 \hline thermal2 & 1.2E6 &  8.6E6 & 16.0s (0.9G) & 4.4s (0.2G) & 34 & 1,582X & 40X\\
 \hline ecology2 & 1.0E6 &   5.0E6 & 12.5s (0.7G) & 3.6s (0.2G) & 47 & 1,728X& 42X \\
 \hline tmt\_sym & 0.7E6 & 5.1E6 & 11.8s (0.6G) & 2.2s (0.1G) & 30 & 796X & 28X\\
 \hline paraboli\_fem & 0.5E6 & 3.7E6 & 6.3s (0.5G) & 1.2s (0.1G) & 25 & 120X & 11X\\
  \hline
\end{tabular}\label{table:uflsdd}
\end{center}
\end{table}
Accurate analysis of on-chip power grids is indispensable for designing modern VLSI chips since it can help reveal critical design issues related to power supply noise, electromigration, etc. However, modern power grid designs can integrate billions of components, which  results in super-linear runtime/memory cost when using direct solution methods. The proposed spectral sparsification technique  allows  developing nearly-linear time iterative solvers for power grid analysis problems. Additionally, even more general transistor-level SPICE-accurate circuit simulations can  potentially benefit  from the proposed spectral graph sparsification algorithm \cite{xueqian:iccad12,xueqian:dac12,lengfei:iccad13}. Table \ref{table:ibmpg} and Table  \ref{table:thupg} demonstrate the DC analysis   results of  IBM and THU power grid design benchmarks \cite{ibmpg,thupg}, showing nearly-linear runtime/memory cost. Similar runtime  scalability is observed from Table \ref{table:uflsdd} for solving sparse matrices from \cite{davis2011matrix}. In all  test cases, the proposed spectral graph sparsification algorithm can find  tree-like ultra-sparsifiers with high spectral similarity. For example, we achieve $\kappa({L_G},{L_P})\approx 16$ ($\sigma\approx 4$), for all IBM power grid test cases, which allows solving the sparse matrices within just a small number of PCG iterations (e.g. $N_I<14$).
%


\subsection{Estimation of Extreme Eigenvalues}
\begin{table}
\begin{center}
 \addtolength{\tabcolsep}{-2.5pt} \centering
\caption{Results of extreme  eigenvalue estimations.}
\begin{tabular}{|c|c|c|c|c|c|c|}
 \hline   Test Cases& $\lambda_{min}$ & $\tilde\lambda_{min}$ &$\epsilon_{\lambda_{min}}$  &  $\lambda_{max}$& $\tilde\lambda_{max}$&$\epsilon_{\lambda_{max}}$ \\
  \hline  fe\_rotor & $1.34$ & $1.40$ &$4.4\%$  &  $120.9$& $116.7$&$3.5\%$ \\
   \hline  pdb1HYS & $1.71$ & $1.89$ &$10.5\%$  &  $120.6$& $113.2$&$6.1\%$ \\
      \hline  bcsstk36 & $1.18$ & $1.27$ &$7.6\%$  &  $96.0$& $92.4$&$3.8\%$ \\
            \hline  brack2 & $1.15$ & $1.20$ &$4.3\%$  &  $92.6$& $90.3$&$2.5\%$ \\
 \hline  raefsky3 & $1.13$ & $1.25$ &$10.5\%$  &  $84.4$& $82.7$&$2.0\%$ \\
   \hline
\end{tabular}\label{table:eigvalue}
\end{center}
\end{table}

In Table \ref{table:eigvalue}, the extreme generalized  eigenvalues ($\tilde\lambda_{min}$ and $\tilde\lambda_{max}$) estimated by the proposed methods (Section \ref{sec_extreme-eigvalue}) are compared with the   ones ($\lambda_{min}$ and $\lambda_{max}$)  computed by the ``\emph{eigs}" function in Matlab for sparse matrices in \cite{davis2011matrix}, while the relative errors ($\epsilon_{\lambda_{min}}$ and $\epsilon_{\lambda_{max}}$) are also shown.  $\tilde\lambda_{max}$ is estimated using less than ten generalized power iterations.  
\subsection{Spectral Ranking of Off-tree Edges}
\begin{figure}
\centering \epsfig{file=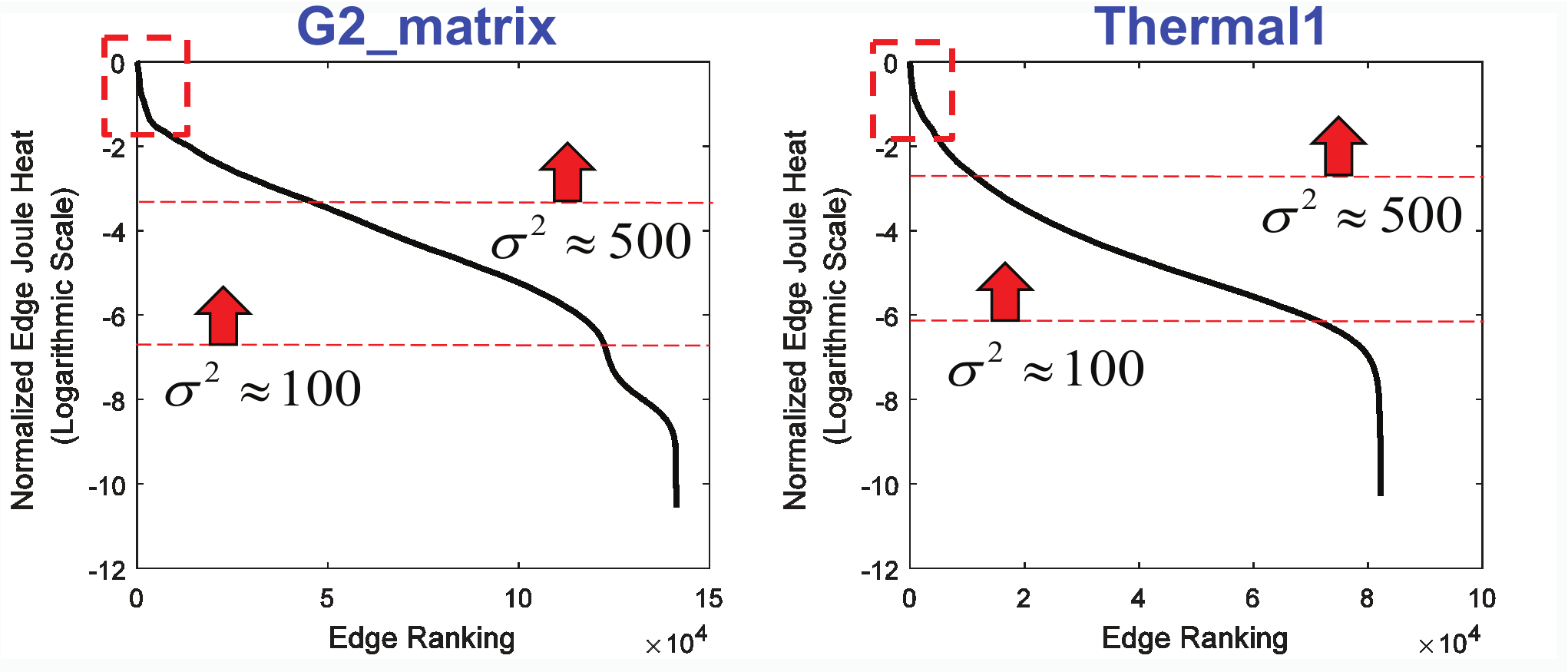, scale=0.43} \caption{Spectral edge  ranking and filtering by normalized Joule heat of off-tree edges for $G2\_circuit$ (left)  and $Thermal1$ (right) test cases\cite{davis2011matrix} with  top off-tree edges highlighted in red rectangles. \protect\label{fig:edge_rank}}
\end{figure}

We   illustrate the results of  spectral edge ranking and filtering according to Joule heat levels computed by  one-step generalized power iteration using (\ref{formula_pwr_iter5new})   in Fig. \ref{fig:edge_rank}  for two  sparse matrices in \cite{davis2011matrix}. The thresholds of normalized edge Joule heat values required for spectral edge filtering are labeled using red dash lines. It is observed in Fig. \ref{fig:edge_rank} there is a sharp change of the top normalized edge Joule heat values, which indicates that there are not  many large eigenvalues of $\mathbf{L_P^{+}L_G}$ in both cases and agrees well with the prior theoretical analysis \cite{spielman2009note}.

\begin{figure*}
\centering \epsfig{file=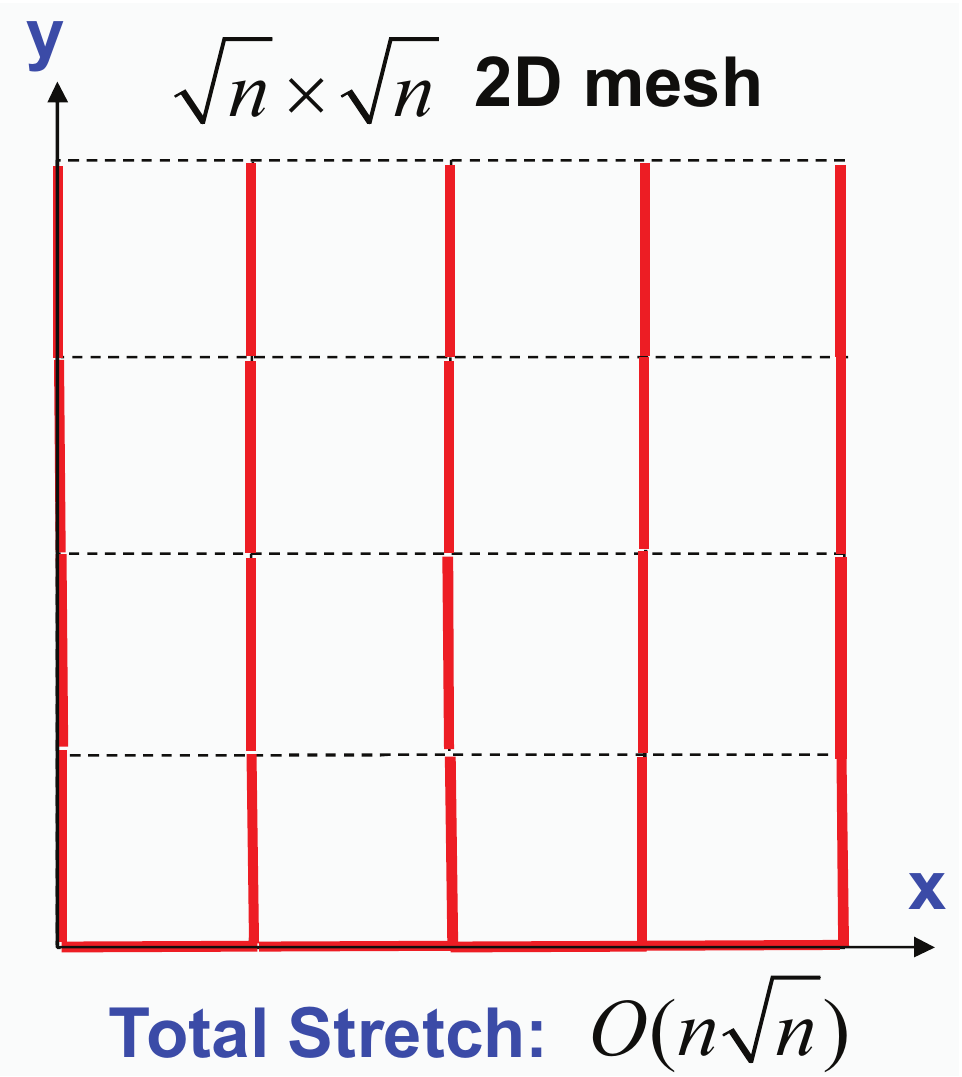, scale=0.475}\epsfig{file=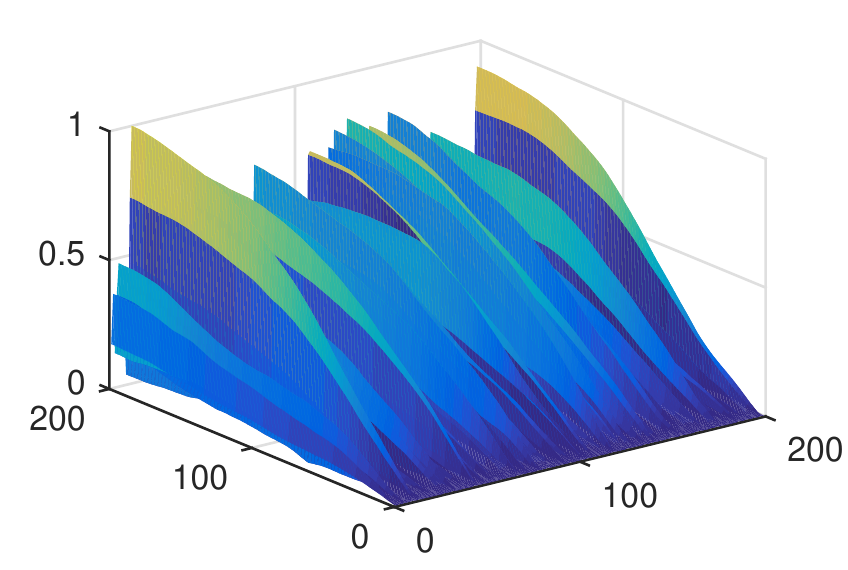, scale=0.765}\epsfig{file=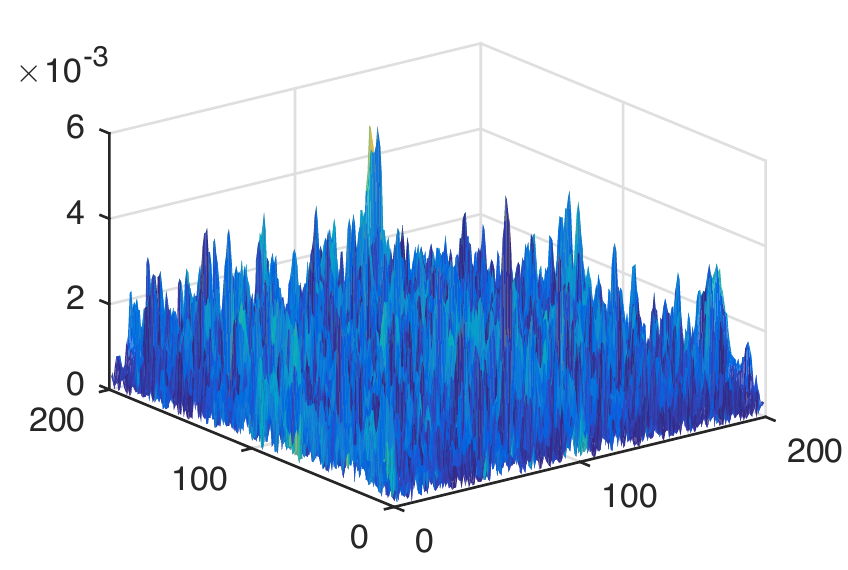, scale=0.765}\caption{Joule heat distributions before (middle) and after (rightmost) adding off-tree edges to the ``hair comb" spanning tree (leftmost). \protect\label{fig:jouleheat_map}}
\end{figure*}
We also demonstrate the results of  Joule heat levels (spectral criticality) of off-tree edges computed by (\ref{formula_pwr_iter5new}) for a random vector using one-step generalized power iteration ($t=1$) for a ``hair-comb" spanning tree shown in Fig. \ref{fig:jouleheat_map}.  It is not difficult to show that  such a spanning tree can not well match the top part of the original 2D grid, so the Joule heat levels  of off-tree edges at the top should be much greater than the ones at the bottom part.  In fact, it has been shown that   the  ``hair-comb"  spanning tree for a $\sqrt{n}\times\sqrt{n}$ 2D mesh will have a total stretch of $\Theta(n \sqrt{n})$ that is mainly contributed by the off-tree edges with largest stretch values near the top part of the mesh grid\cite{peng2013phd}. Using the proposed similarity-aware spectral graph sparsification framework, we are able to efficiently identify and recover the  most spectrally-critical off-tree edges,  thereby dramatically reducing the largest generalized eigenvalues. For example,  the ``hair-comb" spanning tree of a $200\times200$ mesh grid can be dramatically improved in terms of spectral similarity by recovering top $400$   spectrally-critical off-tree edges: $\sigma^2$ is reduced from about $64,000$ to about $100$ ($640\times$ reduction), which can also be indicated by the Joule heat  distributions before and after adding these off-tree edges as illustrated in Fig. \ref{fig:jouleheat_map}.

\subsection{Preservation of Long-Range Effects in the Sparsifier}
\begin{figure*}
\centering \epsfig{file=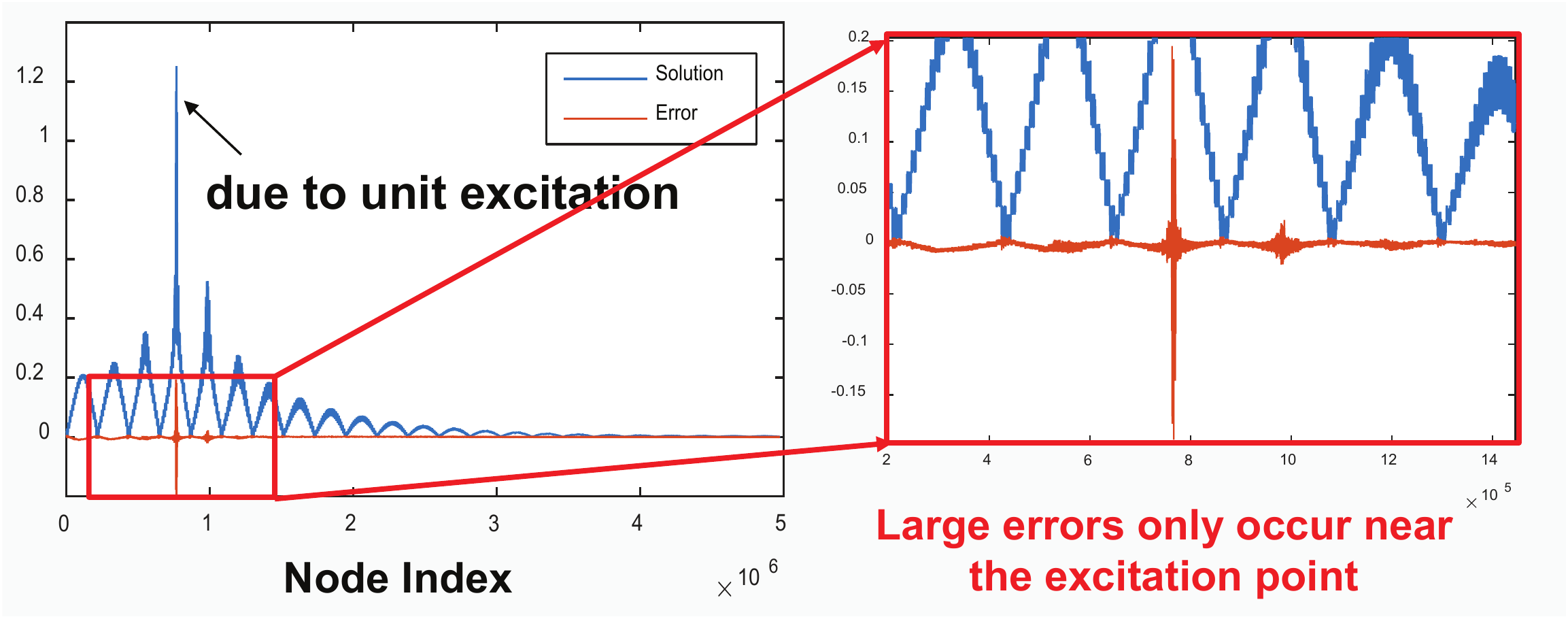, scale=0.608} \caption{Preservation of long range effects in the sparsifier. \protect\label{fig:longrange}}
\end{figure*}

The similarity-aware spectral sparsifier extracted using the proposed framework will  effectively preserve low-frequency graph signals or long-range effects due to the good preservation of graph spectral (structural or global) properties. As an example shown in Fig. \ref{fig:longrange}, the responses of the original on-chip power grid  and its spectrally sparsified grid ($\sigma^2=50$) are obtained by solving $\mathbf{Ax=b}$  and $\mathbf{\tilde{A}\tilde{x}=b}$ respectively, where $\mathbf{A}$ ($\mathbf{\tilde{A}}$) denotes the original (sparsified) conductance matrix and $\mathbf{b}$ denotes a unit excitation right-hand-side (RHS) vector with only a single element being $1$ and others being $0$. Note that we applied a  global edge scaling procedure \cite{zhao:dac19} to the sparsified power grid network in order to match the original node-wise effective resistances in $\mathbf{\tilde{A}}$. If we consider the RHS vector $b$ as the original input graph signal, and the voltage response vector $x$ as the output after graph signal processing, the power grid system can be naturally regarded as a low-pass filter for graph signals. Consequently, the solutions obtained by solving the original and sparsified power grid problems using a unit excitation source can be understood  as the impulse responses of low-pass filters commonly studied in classic Fourier analysis. Fig. \ref{fig:longrange} obviously indicates the good preservation of long range effects or low-frequency components on the  spectral sparsifier obtained using the proposed method.

 \vspace{-0.13cm}
\subsection{An  SDD Matrix Solver with Similarity-Aware Sparsification}
\begin{figure}
\centering \epsfig{file=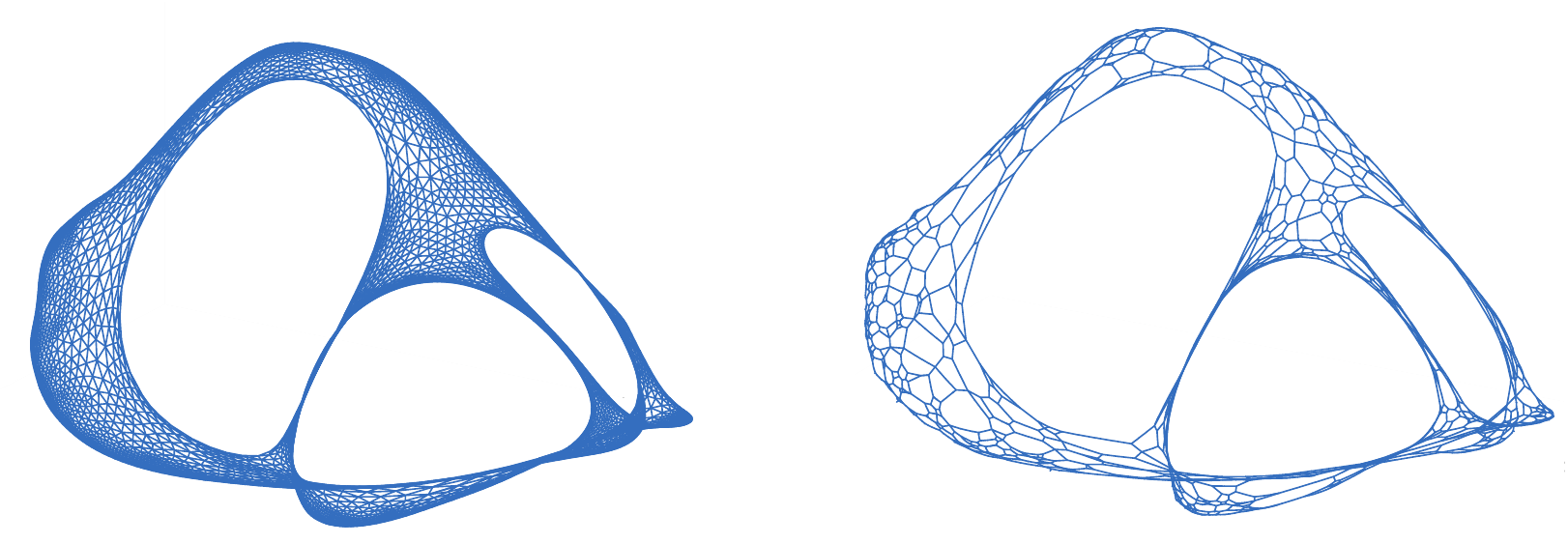, scale=0.535} \caption{Two spectrally-similar airfoil graphs ($\sigma=5$). \protect\label{fig:similarity}}
\end{figure}
Fig.~\ref{fig:similarity} shows the spectral drawings \cite{koren2003spectral} of an airfoil graph \cite{davis2011matrix} as well as its spectrally-similar ($\sigma=5$) subgraph  computed by the proposed similarity-aware spectral sparsification algorithm.
 
The spectral sparsifiers obtained by the proposed similarity-aware algorithm  are also leveraged as  preconditioners in a PCG solver.  The RHS input vector $b$ is generated randomly and the solver is set to converge to  an accuracy level $\mathbf{||A x-b||<10^{-3}||b||}$  for all test cases. In Table \ref{table:filtersdd}, ``$|V|$" and ``$|E|$" denote the numbers of nodes and edges in the original graph, whereas ``$|E_{\sigma^2}|$",  ``$N_{\sigma^2}$" and ``$T_{\sigma^2}$" denote the number of edges in the sparsifier, the number of PCG iterations required for converging to the desired accuracy level, and the total time of graph sparsification for achieving the spectral similarity of $\sigma^2$, respectively. As observed in all test cases, there are very clear trade-offs between the graph density,  computation time,   and spectral similarity for all spectral sparsifiers extracted using the proposed method: sparsifiers with higher spectral similarities (smaller $\sigma^2$)  allow converging to the required solution accuracy level in much fewer PCG iterations, but need to retain more edges in the subgraphs and thus require longer time to compute (sparsify).

\begin{figure}
\centering \epsfig{file=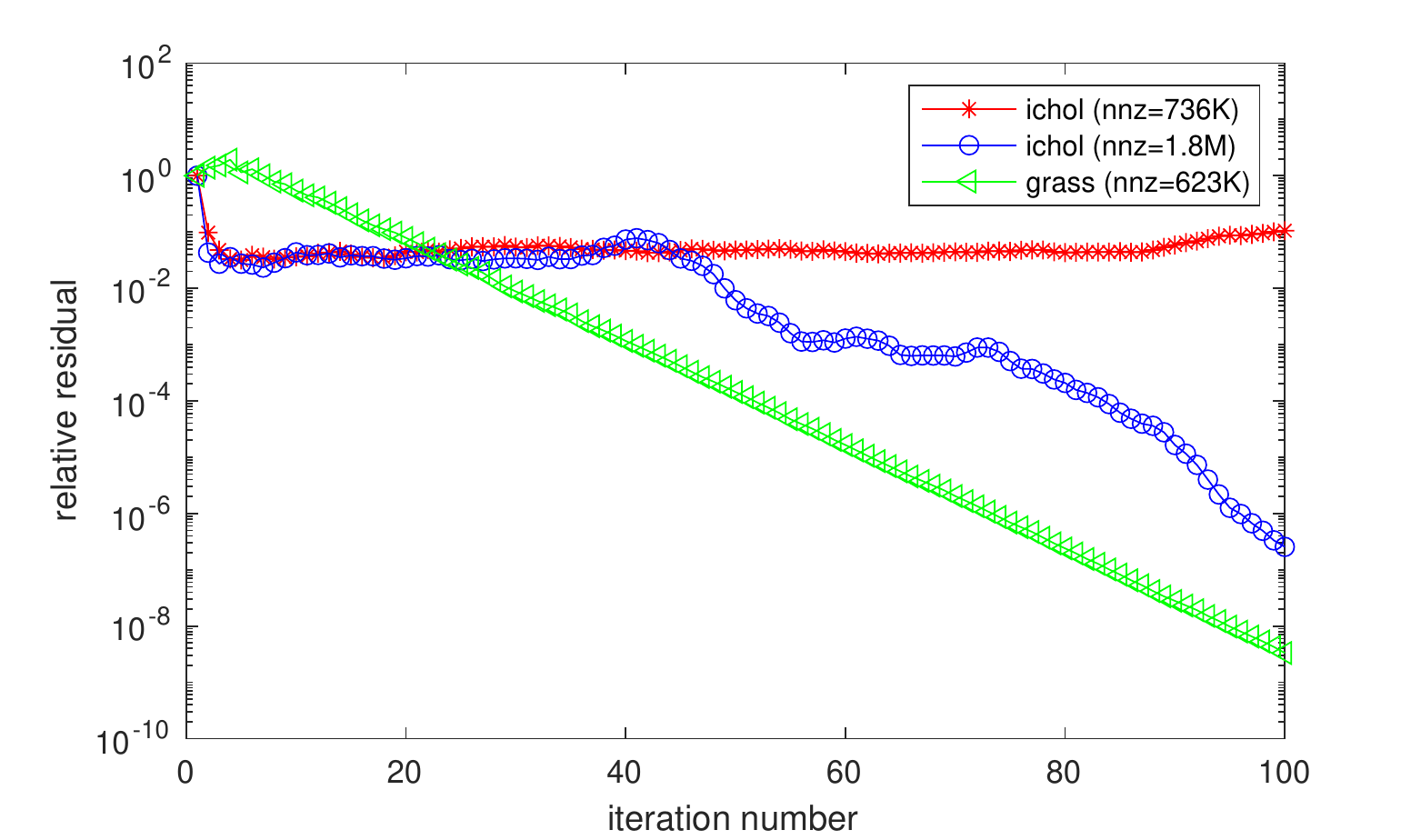, scale=0.56} \caption{PCG convergence results  (G2\_circuit matrix). \protect\label{fig:pcgG2}}
\end{figure}

\begin{table}
\begin{center}
 \addtolength{\tabcolsep}{-2.5pt} \centering
\caption{Results of iterative SDD matrix solver.}
\begin{tabular}{|c|c|c|c|c|c|c|c|c|}
 \hline   Graphs & $|V|$ & $|E|$ & $\frac{|E_{50}|}{|V|}$ & $N_{50}$ & $T_{50}$ &  $\frac{|E_{200}|}{|V|}$ & $N_{200}$ &  $T_{200}$\\
 \hline G3\_circuit & 1.6E6 &  3.0E6 & 1.11 & 21   & 20s & 1.05& 37& 8s \\
 \hline thermal2 & 1.2E6 &  3.7E6 & 1.14  & 20  & 23s & 1.06 & 36 & 9s\\
 \hline ecology2 & 1.0E6 &   2.0E6 & 1.14  & 20  & 16s &  1.06  & 40  & 5s \\
 \hline tmt\_sym & 0.7E6 & 2.2E6 & 1.21  & 19  & 16s &  1.14  & 38  & 4s\\
 \hline paraboli\_fem & 0.5E6 & 1.6E6 &  1.22  & 18  & 16s &  1.09  & 38  & 3s\\
  \hline
\end{tabular}\label{table:filtersdd}
\end{center}
\end{table}
 Similar runtime  scalability is observed from Table \ref{table:filtersdd} for solving sparse matrices from \cite{davis2011matrix}. In all  test cases, the proposed spectral graph sparsification algorithm can find  tree-like ultra-sparsifiers with high spectral similarity. In Fig. \ref{fig:pcgG2}, the relative residual plot (versus PCG iteration number) has been shown  using three preconditioners. As observed, the PCG iterations with the preconditioner obtained by directly running the Cholesky decomposition algorithm for the spectrally-sparsified matrix computed by GRASS converge much faster than the ones with preconditioners generated by using standard incomplete Cholesky decomposition algorithm (with drop tolerances of $0.01$ and $0.001$); it is also observed that the number of non-zeros (nnz) in the preconditioner matrix created by GRASS is the lowest.  
\vspace{-0.101500cm}

\subsection{A Scalable Spectral Graph Partitioner}
It has been shown that by applying only a few inverse power iterations, the approximate Fiedler vector ($u_{f}$) that corresponds to the smallest nonzero eigenvalue  of the (normalized) graph Laplacian matrix can be obtained for obtaining high-quality graph partitioning solution \cite{spielman2014sdd}. Therefore, using the spectral sparsifiers computed by the proposed spectral  sparsification algorithm can immediately accelerate the PCG solver for inverse power iterations, leading to scalable performance for  graph partitioning problems \cite{spielman2014sdd}. In fact, if the spectral  sparsifier is already a good approximation of the original graph, its Fiedler vector can be directly used for   partitioning  the original graph.

We implement the accelerated  spectral graph partitioning algorithm, and test it with sparse matrices in \cite{davis2011matrix} and several 2D mesh graphs synthesized with random edge weights.   As shown in Table \ref{table:fiedler}, the graphs associated with sparse matrices have been partitioned into two pieces using sign cut method \cite{spielmat1996spectral}  according to the approximate Fiedler vectors computed by a few steps of inverse power iterations. The direct solver \cite{cholmod}  and the preconditioned iterative solver are invoked within each inverse power iteration for updating the approximate Fiedler vectors $u_{f}$ and $\tilde u_{f}$, respectively.  $\frac{|V_+|}{|V_-|}$ denotes the ratio of nodes assigned with positive and negative signs according to the approximate Fiedler vector, and ``Rel.Err." denotes the relative error of the proposed solver compared to the direct solver computed by $\frac{|V_{dif}|}{|V|}$, where $|V_{dif}|$ denotes the number of nodes with different signs in $u_{f}$ and $\tilde u_{f}$. `` $T_{D}$" (`` $T_{I}$") and ``$M_{D}$" (`` $M_{I}$") denote   the total solution time (excluding   sparsification time) and   memory cost of  the direct (iterative) method. We extract sparsifiers with $\sigma^2\leq 200$ for all test cases.

\begin{figure}
\centering \epsfig{file=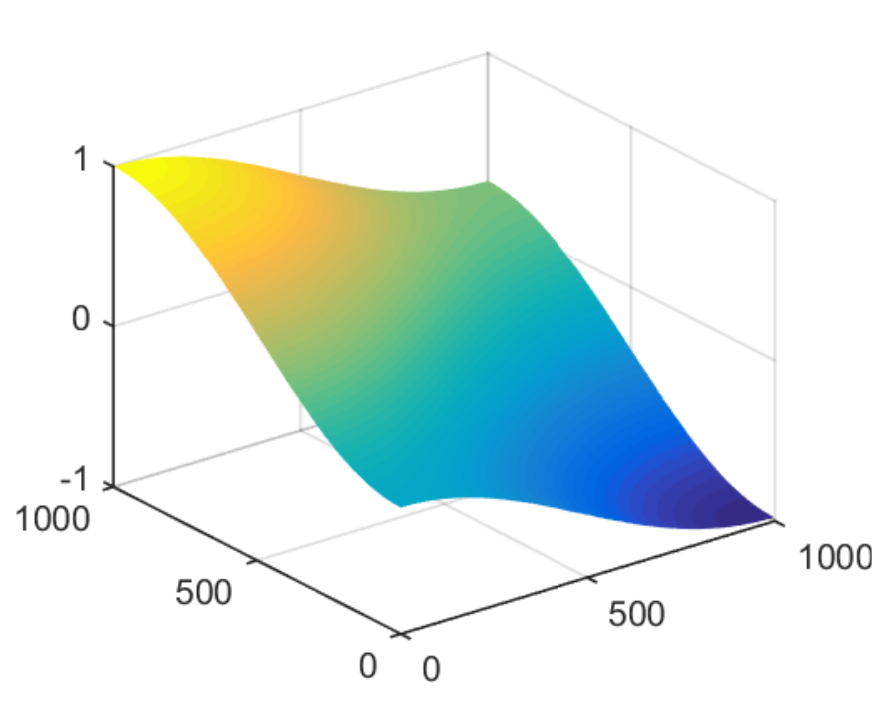, scale=0.485}\epsfig{file=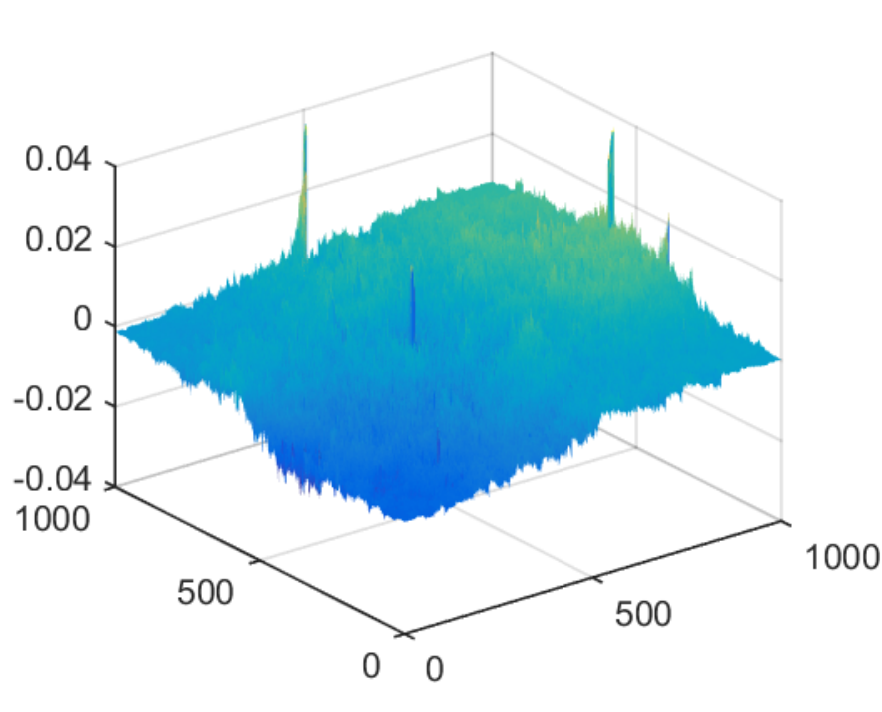, scale=0.485} \caption{The approximate Fiedler vector (left)  and its magnitude error (right) for ``mesh\_1M". \protect\label{fig:fiedlercomp}}
\end{figure}

 It can be observed that the proposed  preconditioned spectral graph partitioner  only results in a very small portion of nodes (0.07\% to 4\%)   assigned with different signs when comparing with the original spectral graph partitioner, while achieving significant runtime and memory  savings ($4$-$10\times$). The approximate Fiedler vector computed by our fast solver for the test case ``mesh\_1M" is also illustrated in Fig. \ref{fig:fiedlercomp}, showing rather good agreement with the true solution.

\begin{table}
\begin{center}
 \addtolength{\tabcolsep}{-0.5pt} \centering
\caption{Results of spectral graph partitioning.}
\begin{tabular}{|c|c|c|c|c|c|}
 \hline   Test Cases & $|V|$ & $\frac{|V_+|}{|V_-|}$ &  $T_{D}$ ($M_{D}$) &  $T_{I}$ ($M_{I}$) & Rel.Err.  \\
  \hline G3\_circuit & 1.6E6 &  1.35 & 52.3s (2.3G) & 7.6s (0.3G) & 2.2E-2  \\
 \hline thermal2 & 1.2E6 &  1.00 & 13.0s (0.9G) & 3.0s (0.2G) & 6.8E-4  \\
 \hline ecology2 & 1.0E6 &   1.03 & 12.1s (0.7G) & 3.4s (0.2G) & 8.9E-3  \\
 \hline tmt\_sym & 0.7E6 & 0.99& 10.2s (0.6G) & 1.9s (0.1G) &2.1E-2  \\
 \hline paraboli\_fem & 0.5E6 & 0.98 & 8.8s (0.4G) & 2.4s (0.1G) & 3.9E-2  \\
  \hline mesh\_1M & 1.0E6 & 1.01 & 10.2s (0.7G) & 1.7s (0.2G) & 3.3E-3  \\
  \hline mesh\_4M & 4.5E6 & 0.99 & 49.6s (3.0G) & 8.2s (0.7G) & 7.5E-3  \\
  \hline mesh\_9M & 9.0E6 & 0.99 & 138.5s (6.9G) & 13.3s (1.5G) & 7.8E-4  \\
  \hline
\end{tabular}\label{table:fiedler}
\end{center}
\end{table}
  \vspace{-0.cm}
\subsection{Sparsification of Other Complex networks}
\begin{table}
\begin{center}
 \addtolength{\tabcolsep}{-0.5pt} \centering
\caption{Results of complex network sparsification.}
\begin{tabular}{|c|c|c|c|c|c|c|}
 \hline   Test Cases & $|V|$ & $|E|$  &  $T_{tot}$& $\frac{|E|}{|Es|}$ & $\frac{\lambda_1}{\tilde{\lambda}_1}$& ${T_{eig}^{o}}({T_{eig}^{s}})$  \\
  \hline fe\_tooth & 7.8E4 & 4.5E5 & 3.0s  & $5\times$ & $8E3$& 14.5s (2.7s)\\
    \hline appu & 1.4E4 & 9.2E5 & 5.4s & $25\times$ & $1E4$&  2,400s (15s)\\
  \hline coAuthorsDBLP & 3.0E5 & 1.0E6 & 7.2s & $3\times$ & $1E3$& 2,047s (36s)\\
\hline auto & 4.5E5 & 3.3E6 &29.0s & $5\times$ & $5E4$&  N/A (54s)\\
  \hline RCV-80NN & 1.9E5 & 1.2E7 & 46.5s  & $36\times$ & $3E4$&  N/A (170s)\\
  \hline
\end{tabular}\label{table:uflsocial}
\end{center}
\end{table}
As shown in Table \ref{table:uflsocial},  a few finite element,   protein, data and social networks  have been  spectrally sparsified to achieve $\sigma^2 \approx 100$ using the proposed similarity-aware method. ``$T_{tot}$" is the total time for extracting the sparsifier, ``$\frac{\lambda_1}{\tilde{\lambda}_1}$" denotes the  ratio of  the largest generalized eigenvalues before and after adding off-tree edges into the spanning tree sparsifier, and  ${T_{eig}^{o}}({T_{eig}^{s}})$ denotes the time for computing the first ten eigenvectors of the original  (sparsified) graph Laplacians using the ``$eigs$" function in Matlab. Since spectral sparsifiers can well approximate the spectral (structural) properties of the original graph, the sparsified graphs can be leveraged for accelerating many numerical and graph-related tasks. For example, spectral clustering (partitioning) using the original ``RCV-80NN"  (80-nearest-neighbor) graph can not be performed on our server with $50 GB$ memory, while it only takes a few minutes  using the sparsified one. 
\subsection{Nearly-linear Runtime Scalability}
\begin{figure}
\centering \epsfig{file=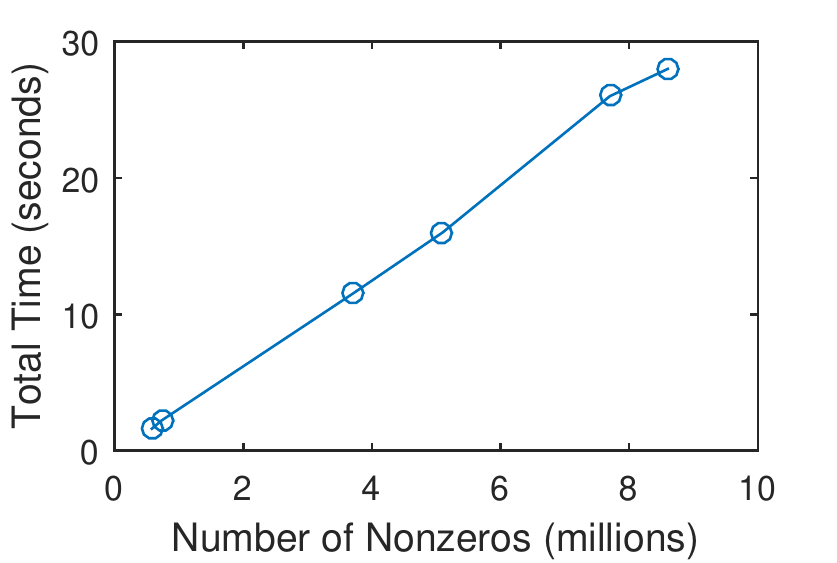, scale=0.8591}\caption{Runtime scalability of the proposed similarity-aware spectral   sparsification approach. \protect\label{fig:runtimeplot}}
\end{figure}

 Our   results show that the proposed method can extract the  similarity-aware spectral sparsifier   in nearly-linear time  as shown in Fig. \ref{fig:runtimeplot}. It should be noted that each spectral sparsifier needs to be extracted once and can be reused or incrementally updated  many times \cite{xueqian:tcad15,lengfei:tcad15}.
\section{Conclusions}\label{conclusion}
This paper introduces a nearly-linear time yet practically efficient    spectral graph sparsification algorithm that can be immediately  leveraged to develop nearly-linear time sparse   matrix solvers and spectral graph (data) partitioning (clustering) algorithms. A novel spectral perturbation based  approach is proposed for constructing an ultra-sparse spectral graph sparsifier by adding the most  spectrally-critical  off-tree edges back to the initial spanning tree subgraph, so that key spectral properties of the original graph  can be very well approximated. Additionally, we also propose  a similarity-aware spectral graph sparsification framework that leverages  efficient spectral off-tree edge embedding and filtering schemes to construct spectral sparsifiers with guaranteed spectral similarity (relative condition number) level. An iterative graph densification  scheme is   introduced to facilitate efficient and effective filtering of off-tree edges  for   highly ill-conditioned problems. Extensive experimental results  show the runtime of the SDD solver and  the spectral graph partitioner scales nearly-linearly with the graph size for a variety of large-scale, real-world  problems, such as VLSI power grid analysis, circuit simulation, finite element problems, transportation and social networks, etc. For instance, a sparse matrix with $40$ million unknowns and $180$ million nonzeros can  be solved within two minutes using a single CPU core and about $6GB$ memory.


\bibliographystyle{abbrv}
{
\bibliography{dac18-spectralgraph}  
}

\begin{IEEEbiography}{Zhuo Feng} (S'03-M'10-SM'13) received the B.Eng. degree
in information engineering from Xi'an Jiaotong University, Xi'an,
China, in 2003, the M.Eng. degree in electrical engineering from
National University of Singapore, Singapore, in 2005, and the Ph.D.
degree in electrical and computer engineering from Texas A\&M
University, College Station, TX, in 2009. He is currently an associate professor at Stevens Institute of Technology. His
research interests include high-performance spectral methods, very large scale integration
(VLSI) and computer-aided design (CAD), scalable hardware and software systems,  as well as heterogeneous parallel computing.

He received a Faculty Early Career Development (CAREER)
Award from the National Science Foundation (NSF) in 2014, a Best
Paper Award from ACM/IEEE Design Automation Conference (DAC) in
2013, and two Best Paper Award Nominations from IEEE/ACM International
Conference on Computer-Aided Design (ICCAD) in 2006 and 2008. He was the principle investigator of the CUDA Research Center  named by Nvidia Corporation. He has
served on the technical program committees of major international conferences
related to electronic design automation (EDA), including DAC, ASP-DAC,
ISQED, and VLSI-DAT, and has been a technical referee for many leading
IEEE/ACM journals in VLSI and parallel computing.  In 2016, he became a co-founder of LeapLinear Solutions to provide highly scalable software solutions for solving sparse matrices and analyzing graphs (networks) with billions of elements, based on the latest breakthroughs in spectral graph theory.
\end{IEEEbiography}

\end{document}